\pdfoutput=1
\newcommand*{\ATLASLATEXPATH}{./}
\documentclass[USenglish,cernpreprint,texlive=2016]{\ATLASLATEXPATH atlasdoc}

\usepackage[subfigure]{\ATLASLATEXPATH atlaspackage}
\usepackage{\ATLASLATEXPATH atlasbiblatex}

\usepackage{\ATLASLATEXPATH atlascontribute}
\usepackage{lineno}
\usepackage{tabularx, multirow, booktabs}
\newcolumntype{Y}{>{\centering\arraybackslash}X}

\usepackage{\ATLASLATEXPATH atlasphysics}

\usepackage{amsmath}
\usepackage{bm}
\usepackage{multirow}
\usepackage{placeins}
\usepackage{\ATLASLATEXPATH vgamma-defs}
\usepackage{graphicx}
\usepackage{caption}
\usepackage{amssymb}

\graphicspath{{figures/}}

\DeclareGraphicsExtensions{   
	    .pdf,.PDF,%
            .png,.PNG,%
		.jpg,.mps,.jpeg,.jbig2,.jb2,.JPG,.JPEG,.JBIG2,.JB2}

\AtlasTitle{Search for heavy resonances decaying to a photon and a hadronically decaying $Z/W/H$ boson in $pp$ collisions at $\sqrt{s}=13$~\TeV\ with the ATLAS detector}

\author{The ATLAS Collaboration}

\AtlasRefCode{EXOT-2016-30}

\PreprintIdNumber{CERN-EP-2018-055}

\AtlasDate{\today}

\AtlasJournal{PRD}
\AtlasJournalRef{Phys. Rev. D 98 (2018) 032015}
\AtlasDOI{10.1103/PhysRevD.98.032015}

\AtlasCoverCommentsDeadline{-}

\AtlasAbstract{Many extensions of the Standard Model predict new resonances
decaying to a $Z$, $W$, or Higgs boson and a photon. This paper presents a
search for such resonances produced in $pp$ collisions
at $\sqrt{s} = 13$~\TeV\ using a dataset with an integrated
luminosity of 36.1~fb$^{-1}$ collected by the ATLAS
detector at the Large Hadron Collider.
The $Z/W/H$ bosons are identified through their decays to hadrons.
The data are found to be consistent with the Standard Model expectation in the entire
investigated mass range. Upper limits are set on the production
cross section times branching fraction for resonance decays to $Z/W+\gamma$ in the mass
range from 1.0 to 6.8~\TeV, and for the first time into $H+\gamma$
in the mass range from 1.0 to 3.0~\TeV.
}

\hypersetup{pdftitle={ATLAS document},pdfauthor={The ATLAS Collaboration}}

\begin{document}

\maketitle

\section{Introduction}
\label{sec:introduction}

Many proposals for physics beyond the Standard Model (SM) include the
prediction of new massive bosons. Examples are Technicolor~\cite{Eichten:2007sx} or Little
Higgs~\cite{ArkaniHamed:2001nc}, as well as extensions to the SM Higgs
sector such as including an additional electroweak singlet scalar~\cite{Low:2011gn}.
Decay modes of these new bosons include
final states with a $Z$ or a $W$ boson and a photon. In addition,
decays of heavy spin-1 bosons to the 125~\GeV{} Higgs boson and a
photon present an interesting search
channel~\cite{Dobrescu:2017sue}. This paper describes a search for
massive neutral and charged bosons decaying to a photon and a $Z$,
$W$, or Higgs boson with subsequent hadronic decay of these
bosons. The search uses 36.1~$\mathrm{fb}^{-1}$ of proton--proton ($pp$)
collision data at a center-of-mass energy \sqs{} = 13~\TeV{} collected
with the ATLAS detector in 2015 and 2016.

The selection of events collected for this search is based on the
presence of high transverse energy photons. The identification of $Z$, $W$, and Higgs
bosons exploits properties of the
highly boosted bosons with merged dijet energy clusters reconstructed as a
large-radius jet. The advantage of this final state is that a large
fraction of events from the heavy resonance decay is detected since the
branching fraction of $Z$ and $W$ bosons into hadrons is approximately 70\%. The
Higgs boson also decays mainly hadronically, dominated by the decay to a $b\bar{b}$ quark pair
with a branching fraction of 58\%. The measurements use the mass of the
large-radius jet and other substructure information to identify the
$Z$, $W$ and Higgs bosons. In addition, for the $Z$ and Higgs bosons,
the sensitivity is increased by identify the long-lived decays of
bottom hadrons within jets.

Previous searches for high-mass resonances decaying to a $Z$ boson and
a photon were conducted by Tevatron's D0
experiment~\cite{Abazov:2008wg} as well as at the Large Hadron
Collider (LHC). The ATLAS experiment performed searches for heavy
resonances decaying to a $Z$ or a $W$ boson and a photon, with
subsequent leptonic decays of the $Z$ and $W$ bosons, using data
collected at $\sqrt{s}$ = 7 and
8~\TeV{}~\cite{STDM-2012-07-witherratum,EXOT-2013-09}, and for heavy resonances
decaying to a $Z$ boson and a photon using 3.2~\ifb{} of $\sqrt{s}$ = 13~\TeV{} data ~\cite{EXOT-2016-02}. In the latter search, both the leptonic and
hadronic decays of the $Z$ boson were used. The ATLAS experiment also
used a 36.1~\ifb{} dataset collected at $\sqrt{s}$ = 13 TeV to search for resonances decaying
to a $Z$ boson and a photon with the decays $Z \to e^+ e^-$, $\mu^+
\mu^-$~\cite{HIGG-2016-14} and $\nu
\bar{\nu}$~\cite{EXOT-2016-32}.
The CMS experiment performed
searches for a heavy resonance decaying to a photon and a hadronically
or leptonically decaying $Z$ boson using datasets
collected at \sqs~=~7,~8, and 13~\TeV{}~\cite{CMS-HIG-13-006,CMS-EXO-16-021,CMS-EXO-16-025,CMS-EXO-17-005}. None of the
searches revealed the presence of a new resonance.

In the present search, scans of invariant mass spectra of the photon--jet system between 1.0 and 6.8 \TeV{} are used to search for narrow signal resonances, denoted generically by $X$ hereafter, decaying to a $Z$, $W$ or Higgs boson and a photon.
A variety of production and decay models are considered in order for the
search to be sensitive to Higgs-like (spin 0), $W$-like (spin 1), and graviton-like (spin 2) bosons~\cite{Artoisenet:2013puc, Pappadopulo:2014qza}.
Gluon--gluon fusion production is considered for spin-0 and spin-2
resonances decaying to $Z\gamma$~\cite{Artoisenet:2013puc}. Production
via quark--antiquark annihilation is considered for a spin-2
resonance decaying to $Z\gamma$~\cite{Artoisenet:2013puc}, a spin-1 resonance decaying to
$W\gamma$~\cite{Pappadopulo:2014qza}, and for the first time a spin-1 $H\gamma$ resonance.

\section{The ATLAS detector}
\label{sec:detector}
The ATLAS detector~\cite{PERF-2007-01} consists of a tracking detector within a superconducting solenoid providing a \SI{2}{\tesla} axial magnetic field, electromagnetic and hadronic calorimeters, and a muon spectrometer incorporating three large superconducting toroid magnets. The inner detector, consisting of silicon pixel, silicon microstrip and transition radiation tracking detectors, covers the pseudorapidity\footnote{
ATLAS uses a right-handed coordinate system with its origin at the nominal interaction point (IP) in the center of the detector and the $z$-axis along the beam pipe. The $x$-axis points from the IP to the center of the LHC ring, and the $y$-axis points upward. Cylindrical coordinates $(r,\phi)$ are used in the transverse plane, $\phi$ being the azimuthal angle around the $z$-axis. The pseudorapidity is defined in terms of the polar angle $\theta$ as $\eta=-\ln\tan(\theta/2)$.} range $|\eta| < 2.5$.
The energies of photons and jets are measured primarily by the calorimeter system, which covers the pseudorapidity range $|\eta| < 4.9$.
The electromagnetic calorimeter is a high-granularity liquid-argon (LAr) sampling calorimeter with lead absorber plates and is located just outside the solenoid.
It spans the region $|\eta|< 3.2$ with barrel and endcap sections, segmented into three layers longitudinal in shower--depth in the region $|\eta|< 2.5$, with  \(\Delta\eta \times \Delta\phi\) readout granularity in the second layer of \(0.025 \times 0.025\).
Beyond the cryostat of the electromagnetic calorimeter a steel/scintillator tile hadronic calorimeter covers the region  $|\eta| < 1.7$. It is also segmented into three layers longitudinal in shower--depth  with lateral readout granularity of \(0.1 \times 0.1\) in \(\Delta\eta \times \Delta\phi\).
Two copper/LAr hadronic endcap calorimeters with similar granularity cover the region $1.5 < |\eta|< 3.2$. In the forward region, electromagnetic and hadronic energy measurements are provided by copper/LAr and tungsten/LAr modules, respectively.
A two-level trigger system~\cite{TRIG-2016-01} is used to select the events to be recorded.
The first level of the trigger is implemented in hardware using a subset of the detector information to reduce the event rate to at most 100 kHz from the beam bunch crossing rate of 40 MHz.
The final data selection is done with a software-based trigger that reduces the event rate to an average of 1 kHz.

\section{Data and Monte Carlo samples}
\label{sec:data}

The data used in this search were collected in 2015 and 2016 from LHC $pp$ collisions with a 13~\TeV{} center-of-mass energy and 25~ns bunch spacing.
In these collisions, the average number of inelastic $pp$ collisions in each bunch crossing (referred to as pileup) was approximately 25.
Events were recorded using a single-photon trigger that imposed a transverse energy threshold of 140~\GeV{} and loose photon identification requirements based on cluster shower-shape variables~\cite{PERF-2013-04}.
The photon trigger is fully efficient for the events selected for this analysis. After requiring all ATLAS subdetectors to be operational, the resulting integrated luminosity is 36.1~\ifb.


Simulated signal events are used to optimize the event identification and estimate the efficiency of the event reconstruction and selection.
SM background processes were simulated to test the parameterization of the jet--photon invariant mass spectra, which is used in the data-driven estimation of the background.
All simulated signal and background event samples were generated with Monte Carlo (MC) techniques as described below.

The production and decay of spin-0 and spin-2 \( Z\gamma \) resonances, spin-1 \(W \gamma\) resonances, and spin-1  \(H \gamma\) resonances were modeled assuming a narrow-width approximation.
The decay width of each resonance was set to 4 \MeV{}, which is much smaller than the experimental resolution, and interference between these resonant processes and non-resonant SM production of the corresponding final states was neglected.

The \( Z\gamma \) scalar resonances produced via  gluon--gluon fusion, \(gg  \to X \to Z\gamma\), were modeled by \POWHEGBOX~\cite{Alioli:2010xd,Bagnaschi:2011tu},
with the CT10 parton distribution function~(PDF) set~\cite{Lai:2010vv},
interfaced with \PYTHIAV{8.186}~\cite{Sjostrand:2007gs} for the underlying event, parton showering and hadronization with the CTEQ6L1 PDF set~\cite{Pumplin:2002vw} and a set of tuned underlying-event parameters called the AZNLO tune~\cite{STDM-2012-23}. This model
is the same as that used for SM $H \to Z\gamma$ production with the resonance mass
varied but the width held fixed.


The  spin-2 \(gg  \to X \to Z\gamma\) and \(q\bar{q} \to X \to Z\gamma\), and the spin-1 \(q\bar{q} \to X \to W\gamma\) and  \(q\bar{q} \to X \to H\gamma\) resonant processes were modeled via effective couplings implemented in \MGaMCatNLOV{v2.3.3}~\cite{Alwall:2011uj},
interfaced to the \PYTHIAV{8.186} parton shower model with the A14 parameter tune~\cite{ATL-PHYS-PUB-2014-021} and the NNPDF2.3LO PDF set \cite{Ball:2012cx}. 
The \( Z\gamma \) models produce transversely polarized $Z$ bosons and the \( W\gamma \) models produce longitudinally polarized \(W\) bosons. 
In these samples, the \(W\) and \(Z\) bosons are forced to decay hadronically, and the Higgs boson is forced to decay to \( b\bar{b} \).

The dominant SM backgrounds are prompt photons produced in association with jets, hadronically decaying \(W\) or \(Z\) bosons produced in association with a photon, and \(t\bar{t} + \gamma\) events.
The samples of events containing a photon with associated jets were simulated using the \SHERPAV{2.1.1} generator~\cite{Gleisberg:2008ta}, requiring a photon transverse energy above 35~GeV.
Matrix elements were calculated at leading order with up to either three or four partons, depending on the transverse energy of photon.
They were then merged with the \SHERPA parton shower~\cite{Schumann:2007mg} using the ME+PS@LO prescription~\cite{Hoeche:2009rj}.
\SHERPAV{2.1.1} was also used to simulate events containing an on-shell, hadronically decaying \(W\) or \(Z\) boson and a photon.
Matrix elements were calculated with up to three additional partons at leading order using the \textsc{Comix}~\cite{Gleisberg:2008fv} and \textsc{OpenLoops}~\cite{Cascioli:2011va} matrix element generators, and merged with the \SHERPA parton shower using the ME+PS@NLO prescription~\cite{Hoeche:2012yf}.
For these \(\gamma+\)jet and \(W/Z +\) $\gamma$ simulations, the CT10 PDF set was used in conjunction with a dedicated parton shower tuning developed by the authors of \SHERPA. 
The \( t\bar{t} + \gamma\) events were modeled using \MGaMCatNLOV{v2.3.3} interfaced to the \PYTHIAV{8.186} parton shower with the A14 parameter tune and the NNPDF2.3LO PDF set.
In all MC samples, EvtGen v1.2.0~\cite{EvtGen} was used to model charm and beauty hadron decays.

The simulated signal and SM background events were processed by a detailed \GEANT~4~\cite{Agostinelli:2002hh} simulation of the ATLAS detector~\cite{SOFT-2010-01}.
In all simulated signal and background samples, the effects of overlapping inelastic $pp$ collisions were included by overlaying multiple events simulated with \PYTHIAV{8.186} using the A3 set of tuned parameters~\cite{ATL-PHYS-PUB-2012-003} and the MSTW2008LO PDF set~\cite{Martin:2009iq}.
The simulated events were weighted so that the distribution of the number of pileup interactions in the simulation matched the one in the data. 
The simulated events were then passed through the same event reconstruction algorithms used for the data, including corrections for known differences between data and simulation in the efficiencies of photon reconstruction and selection, in the photon and jet energy scale and resolution, and in the tagging efficiency of heavy-flavor jets.

\section{Event selection}
\label{sec:selection}

\subsection{Reconstruction of photons and jets}

The photon reconstruction is described in Ref.~\cite{ATL-PHYS-PUB-2016-014}.
A discriminant, based on lateral and longitudinal shower profiles, is constructed to distinguish prompt photons from hadrons as well as photons from decays of mesons inside jets.
In this analysis, two levels of selections are applied: the \textit{Loose} selection criterion defined in Ref.~\cite{ATL-PHYS-PUB-2016-014} at the trigger level, and the \textit{Tight} selection criterion for the final analysis selection.
In the final selection, photon candidates are required to have transverse energy above 250~\GeV{} and
pseudorapidity $|\eta|<$ 1.37. These criteria are applied to reduce the contribution from the SM production of prompt photons with associated jets.
The efficiency of a photon within that region to pass the \textit{Tight} selection criterion is above 90\%.
The transverse energy~\etiso{}, deposited within a cone of size $\Delta R \equiv\sqrt{(\Delta \eta)^{2}+(\Delta \phi)^{2}} =$~0.4 around the photon cluster, is corrected for the photon energy deposited outside of the cluster, underlying event, and multiple $pp$ interactions; it is required to satisfy $\etiso < 2.45~\GeV{} + 0.022 \ E_{\mathrm{T}}^{\gamma}$, where $E_{\mathrm{T}}^{\gamma}$ is the transverse energy of the photon cluster. The requirement is applied to reduce contamination from hadrons misidentified as photons and is 98\% efficient for prompt photons passing the \textit{Tight} criterion.  
Corrections mitigating the differences between the calorimeter response to photons in data and simulation are derived using collision data. The photon energy calibration is derived using samples of simulated events followed by data-driven corrections~\cite{ATL-PHYS-PUB-2016-015}.

Large-radius (large-$R$) jets are reconstructed from topological energy clusters (topocluster)~\cite{PERF-2014-07} in the calorimeter, using the \antikt algorithm~\cite{antikt} with a radius parameter $R$ = 1.0.
To reduce contributions to the jet transverse momentum (\pt{}) and mass arising from pileup, a trimming procedure~\cite{Krohn:2009th} is applied where subjets with $R$ = 0.2 and carrying less than 5\% of the original \pt{} of the jet are removed. Jets are calibrated to the level of stable final-state particles using simulation~\cite{PERF-2016-04}. Differences between data and simulation in jet energy scale and resolution are corrected using in situ methods~\cite{ATLAS-CONF-2017-063}. The jet energy resolution for jets with \pt{} of 1~\TeV{} is approximately 5\%. Jet candidates are required to have \pt{} above 200~\GeV{} and $|\eta|<$ 2.0 to ensure tracking detector coverage within the jet cone.
The jet candidates are required to be separated from any photon candidate by $\Delta R > 1.0$.

The jet mass is computed as a weighted combination of \textit{calorimetric mass}
and \textit{track}-\textit{assisted mass}~\cite{ATLAS-CONF-2016-035}. The calorimetric
mass is computed from the massless topocluster four-momenta. The track-assisted
mass incorporates information from the calorimeter and the four-momenta of tracks, which are matched to the jets using
ghost association~\cite{Cacciari:2007fd} and are matched to the primary vertex, which is defined as the vertex with largest sum of squared momenta of associated tracks in the event~\cite{PERF-2015-01}.
The relative weighting of the calorimetric and track-assisted mass in the final mass calculation
is based on the expected resolutions of the two mass variables. The mass
resolution in the peak of the jet mass distribution for jets originating
from $Z$, $W$ or Higgs bosons ranges from 7\% to 15\% for jets
with transverse momenta of 500~\GeV{} to 2500~\GeV{}.
Reconstructed jet mass
distributions for various signal hypotheses are shown in
Figure~\ref{fig:mass-sig}. Peaks centered at the $Z$, $W$, and Higgs
boson masses are clearly visible. In the mass distributions of jets
arising from the $Z$ and Higgs boson decays, a feature at low jet
invariant mass is also prominent for decays of resonances with
masses of 2~\TeV{} or lower. This is a result of energy flow outside of
the jet cone. For the $W\gamma$ spin-1 resonance decays the effect is
mitigated due to the longitudinal polarization of the $W$ boson,
which enhances the collimation of the decay products.
The lower mass side of the cores of the $Z$ boson and Higgs boson mass peaks are enhanced from loss of neutrino energy in the jet when at least one of the $b$-hadrons decays semileptonically.

The jet mass and the \decr{} jet substructure discriminant are employed to distinguish
between jets originating from hadronically decaying $Z$ or $W$
bosons and jets originating from quarks and gluons.
The variable \decr{} is defined as the ratio of two-point and three-point energy
correlation functions~\cite{Larkoski:2014gra,Larkoski:2013eya}, which
are based on the energies and pairwise angular distances of particles
within a jet. The performance of this discriminant has been studied in MC simulations and
data~\cite{ATL-PHYS-PUB-2015-033,ATLAS-CONF-2015-035}.
Upper and lower bounds on the jet mass and upper bound on \decr{} are
tuned to achieve 50\% efficiency for jets with \pt{} in the range of 300~\GeV{} to 2500~\GeV{} from decays of a $Z$ or $W$ boson.
The fraction of jets originating from a quark or a gluon passing this selection varies between approximately 2.2\% and 1.3\% in this \pt{} range.
The \decr{} discriminant is not used for the $H\rightarrow b\bar{b}$ selection.

\begin{figure}
\begin{center}
\subfigure[]{\includegraphics[width=0.49\textwidth]{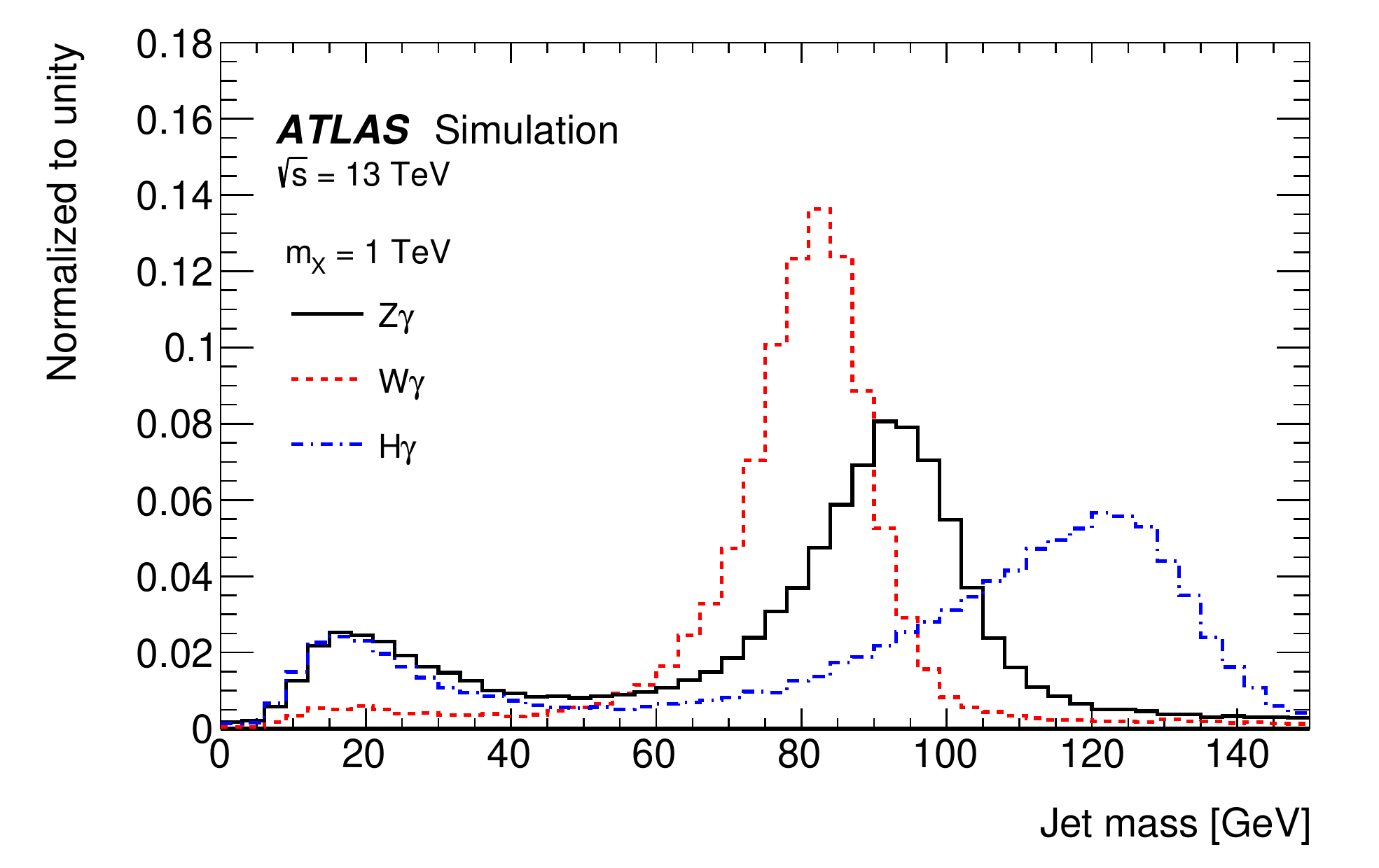}}
\subfigure[]{\includegraphics[width=0.49\textwidth]{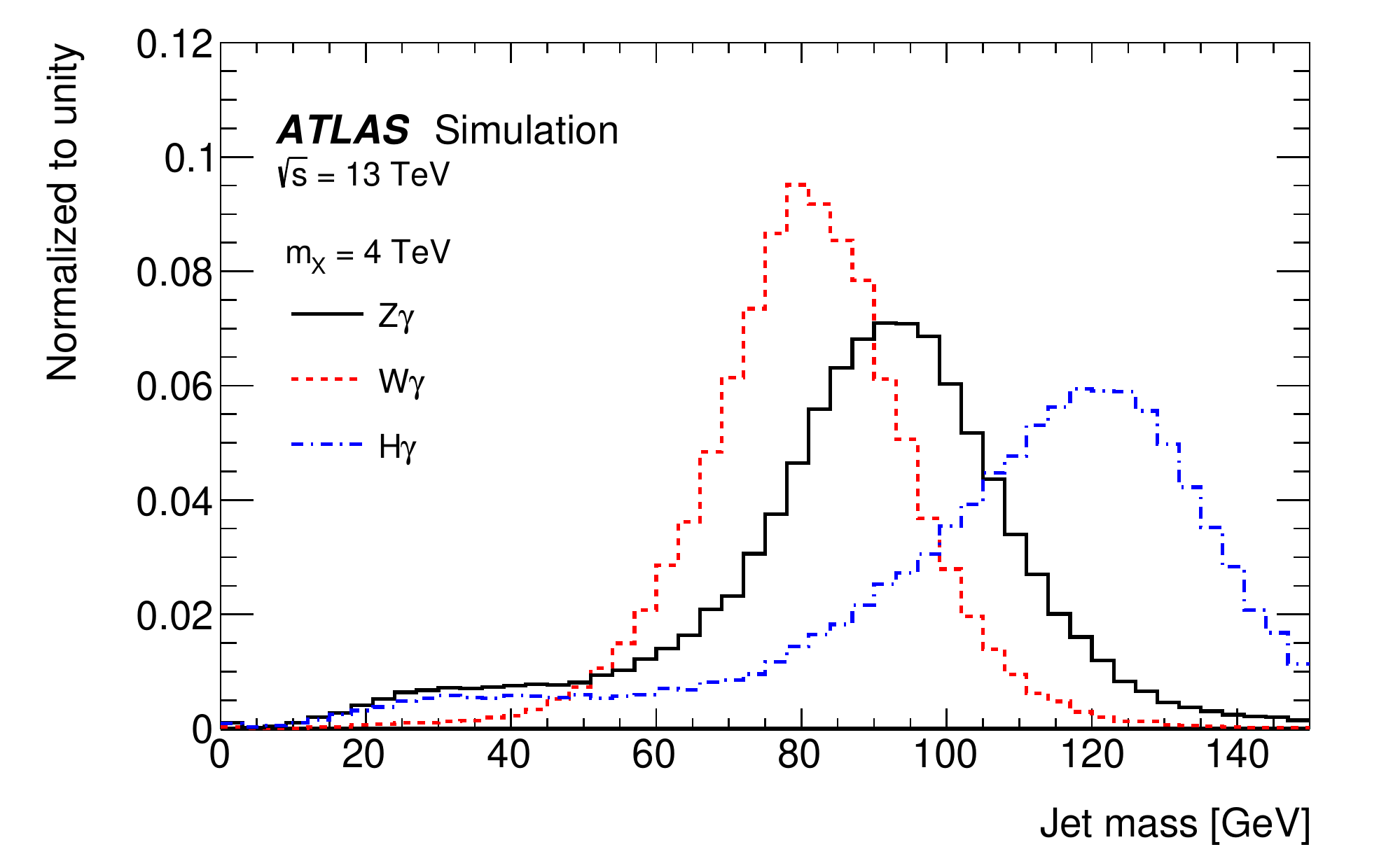}}
\caption{The mass distributions of large-$R$ jets originating from $Z$, $W$, and $H$ bosons resulting from decays of a
  resonance with a mass (a) $m_{X}$ = 1~\TeV{} and (b) $m_{X}$ = 4~\TeV{} in samples of simulated events.
  Only the hadronic $Z$, $W$ and $H\rightarrow b\bar{b}$ decays are considered in these figures.
  Only the spin-0 model is shown for the $Z\gamma$ decay as the jet mass distribution is not sensitive to the signal spin hypothesis.}
\label{fig:mass-sig}
\end{center}
\end{figure}

Track-jets are reconstructed using the \antikt algorithm with radius parameter $R$ = 0.2 and tracks matched to the large-$R$ jets using  ghost association.
Track-jets that contain $b$-hadrons are identified using the MV2c10 tagging algorithm,
which exploits the lifetime of $b$-hadrons and the kinematic properties of their charged decay products~\cite{PERF-2012-04,ATL-PHYS-PUB-2016-012}.
The efficiency of the algorithm is 70\% when applied to $b$-jets in simulated $t\bar{t}$ events.
The fraction of track-jets originating from a light quark or a gluon, tagged as originating from a $b$-hadron, is approximately 0.8\% in simulated $t\bar{t}$ events. 

\subsection{Event selection and categorization}

Events considered for analysis are triggered by the presence of a photon candidate with \pt{} greater than 140~\GeV{}.
A small fraction of events where adverse instrumental effects were
identified are removed. The baseline selection identifies events with
one photon candidate and one large-$R$ jet candidate.
If more than one photon or jet candidate are found, only the highest-\pt{} objects are considered further. The efficiency of the selection
for a resonance with a mass of 3~\TeV{} varies from approximately
60\% to 80\% depending on the signal hypothesis, as shown in
Figure~\ref{fig:sigEffBase}. The decrease of the baseline selection
efficiency for resonances with lower mass is due to the kinematic
thresholds of 200~\GeV{} and 250~\GeV{} required for the jet and photon \pt{}
respectively. Differences in the baseline efficiencies between the
resonance types are due to different angular distributions of the
produced photon--jet system (depending on the spin hypothesis and production mode)
and therefore different probabilities to pass the photon and jet \pt{} and $|\eta|$ requirements.

\begin{figure}
\begin{center}
  \includegraphics[width=0.85\textwidth]{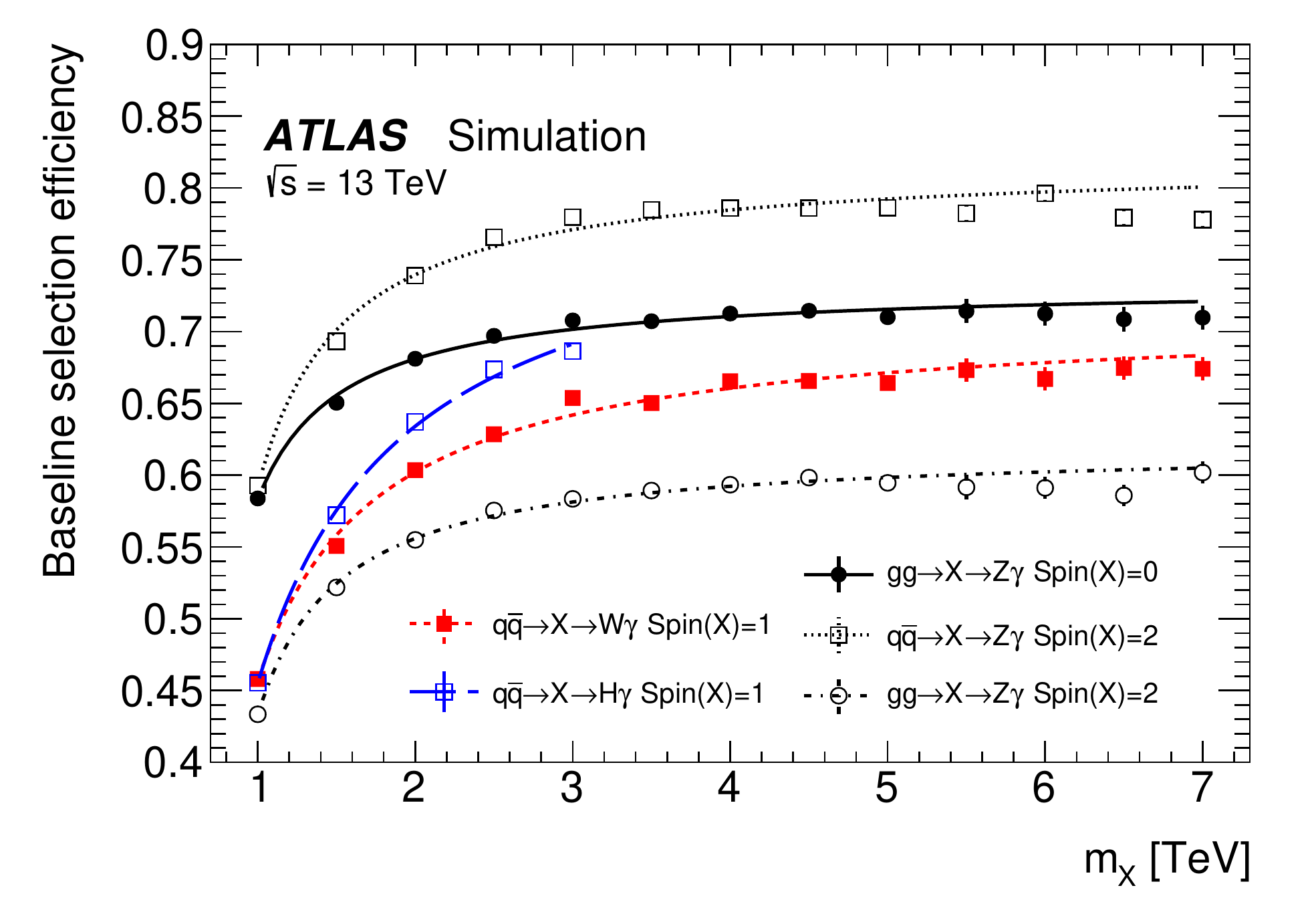}
\end{center}
\caption{Efficiencies for signal events to pass the baseline selection
  as a function of the resonance mass $m_{X}$ for different
  signal models, production modes, and spin hypotheses. The uncertainty bars
  shown are statistical only. The lines represent polynomial fits to the simulated data points. The line corresponding to the $H\gamma$ resonance baseline selection efficiency is discontinued above $m_{X}$ of 3 \TeV{} since the search for $H\gamma$ does not cover this resonance mass region.} 
\label{fig:sigEffBase}
\end{figure}

Following the baseline selection, events are classified into four or fewer
subsamples to improve the expected signal sensitivity. The
categorization is made in order of decreasing background
rejection to achieve high sensitivity throughout the resonance mass
($m_{X}$) range. For resonance masses below 3~\TeV{} it is
desirable to maximally suppress the SM background, while for very high
$m_{X}$ values, due to the steeply falling jet--photon invariant mass (\mgj{}) distribution of background processes, a
loose selection is appropriate. For the $Z\gamma$ search the
categories, defined in the next paragraph, are: BTAG, D2, VMASS, and ELSE. In the $W\gamma$ search
only the D2, VMASS, and ELSE categories are used. The $H\gamma$
search employs only the BTAG category. The process of event
categorization proceeds sequentially starting from the events found in
the baseline selection as shown in the diagrams in
Figure~\ref{fig:cat_flow}.

The first subsample in the $Z\gamma$ and $H\gamma$ searches, the
BTAG category, captures events where the two leading track-jets associated
with the selected large-$R$ jet candidate are $b$-tagged, exploiting
the decays of $Z$ and $H$ bosons to a $b\bar{b}$ quark pair together with strong background suppression.
A window requirement on the jet mass is also applied. In the $Z\gamma$ search the mass
interval grows from 80$-$106~\GeV{} for jets with
\pt{} of 500~\GeV{} to 70$-$110~\GeV{} for jets with \pt
of 2.5~\TeV{}.
These mass intervals are varied such that an
approximately constant signal efficiency across the whole jet \pt{} spectrum is maintained,
accounting for the jet mass resolution increase as a
function of jet \pt{}. In the $H\gamma$ search the mass interval is
93$-$134~\GeV{} independently of the jet \pt{}. In the $Z\gamma$ and $H\gamma$ searches the jet
must also have fewer than 30 associated tracks ($n_{\mathrm{trk}}$) originating from the primary vertex
for the event to be accepted in this category. This requirement is
made to reject gluon-initiated jets that mimic a two-subjet structure due
to gluon splitting~\cite{EXOT-2016-19}. Relative to the baseline selection, the efficiency of selecting in the BTAG category an event originating from
a $Z\gamma$ resonance with a mass of 1~\TeV{}, in the hadronic $Z$ boson decay mode, is 3\% to 4\% depending on the spin hypothesis and
production mode, while only 0.02\% of background events enter this category for the same mass value.
For the $H\gamma$ resonances with a mass of 1~\TeV{}, in the $H\rightarrow b\bar{b}$ decay mode, the selection efficiency is 25\%.
As shown in Figure~\ref{fig:cat_eff}, the BTAG category becomes ineffective for capturing signal events originating from resonances with masses higher than approximately 3~\TeV{} due to $b$-tagging inefficiency for highly boosted jets.
In the $H\gamma$ search the categorization process is stopped at this stage, with the remaining events in the baseline selection being rejected from further analysis.

Events not entering the BTAG category in the $Z\gamma$ search and events from the baseline set in the $W\gamma$ search are placed in the D2 category if the selected jet satisfies the combined jet mass and \decr{} discriminant requirements.
In the $W\gamma$ search the $n_{\mathrm{trk}}$~<~30 requirement must also be satisfied. The mass window requirement in the $Z\gamma$ search is the same as in the BTAG category. In the $W\gamma$ search the mass
interval grows from 71$-$95~\GeV{} for jets with
\pt{} of 500~\GeV{} to 60$-$100~\GeV{} for jets with \pt
of 2.5~\TeV{}.
Approximately 20\% to 25\% (depending on the spin hypothesis and the production mode) of 1~\TeV{} $Z\gamma$ resonance decays passing the baseline selection enter this category with the fraction increasing to 22\%$-$28\% for 4~\TeV{} resonances.
For the $W\gamma$ decays the fraction is approximately 40\% for $m_{X}$ below 4~\TeV{}.
The difference in the D2 categorization efficiency between the $Z\gamma$ and $W\gamma$ processes is explained by differences between the angular distributions of the $Z$ and $W$ boson decay products due to the longitudinal polarization of the $W$ boson and transverse polarization of the $Z$ boson.
For both the $Z\gamma$ and $W\gamma$ resonances the D2 categorization efficiency decreases for $m_{X}$ above 4~\TeV{}, as shown in Figure~\ref{fig:cat_eff},  due to the falling discriminating power of the \decr{} variable. Only approximately 1\% (0.5\%) of background events passing the baseline selection for resonance masses of 1~\TeV{} (4~\TeV{}) enter this category.

Events that fail to enter either of the first two categories and only pass the jet mass selection enter the VMASS category. The efficiency to enter this category, relative to the baseline selection, is 24\% to 27\% for $Z\gamma$ events from a resonance at 1~\TeV{}, growing to 35$-$36\% for a resonance at 4~\TeV{}, and approximately 30\% for $W\gamma$ events. Approximately 9\% of the background events enter this category after passing the baseline selection.

Finally, the remaining events passing the baseline selection are assigned to the ELSE category. The $Z\gamma$ resonance events enter this category with an efficiency of 40\% to 50\%, and $W\gamma$ events enter with approximately 30\% efficiency for resonances below 4~\TeV{}, growing to 50\% for a resonance at 7~\TeV{}.

\begin{figure}
\begin{center}
  \subfigure[]{\includegraphics[width=0.48\columnwidth]{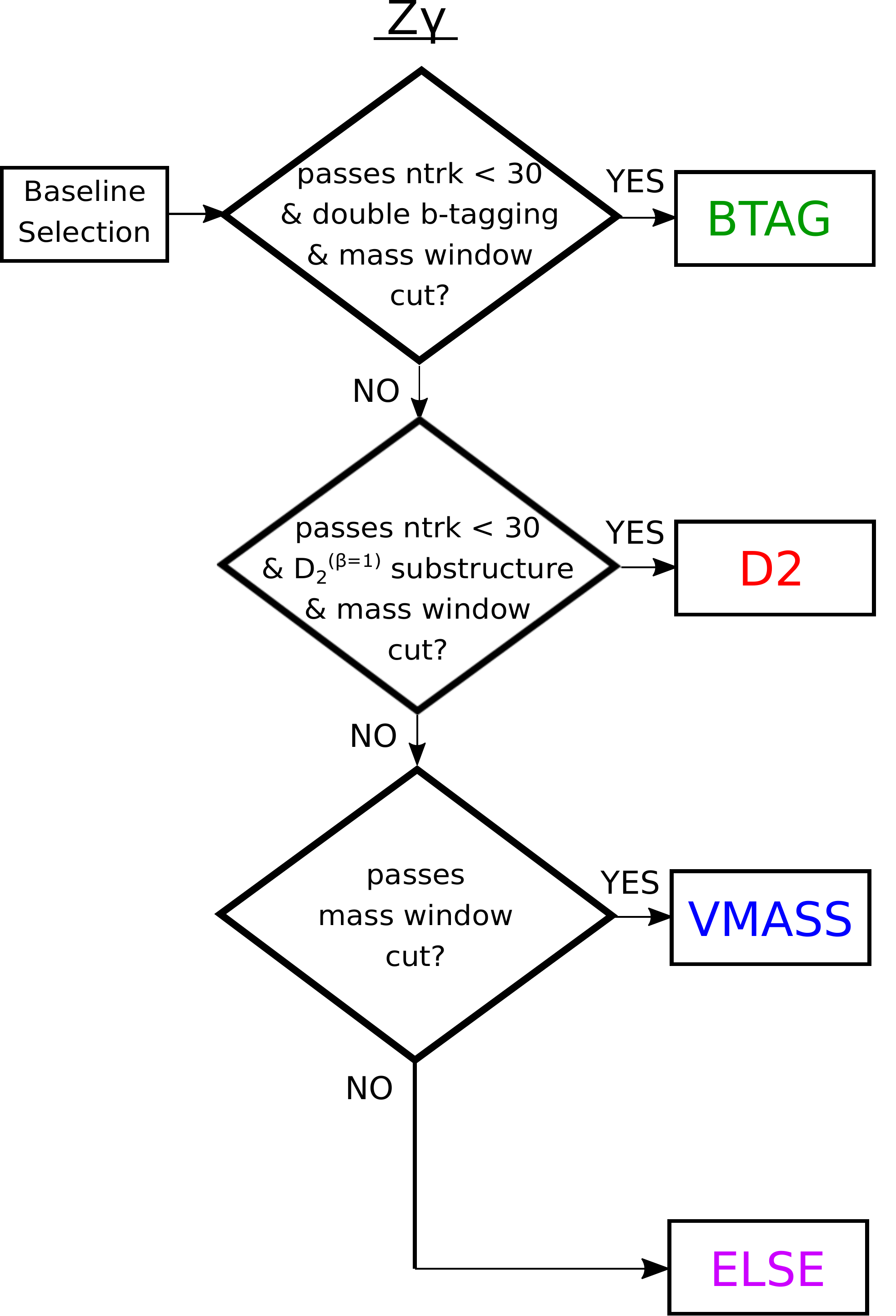}}
   \subfigure[]{\includegraphics[width=0.48\columnwidth]{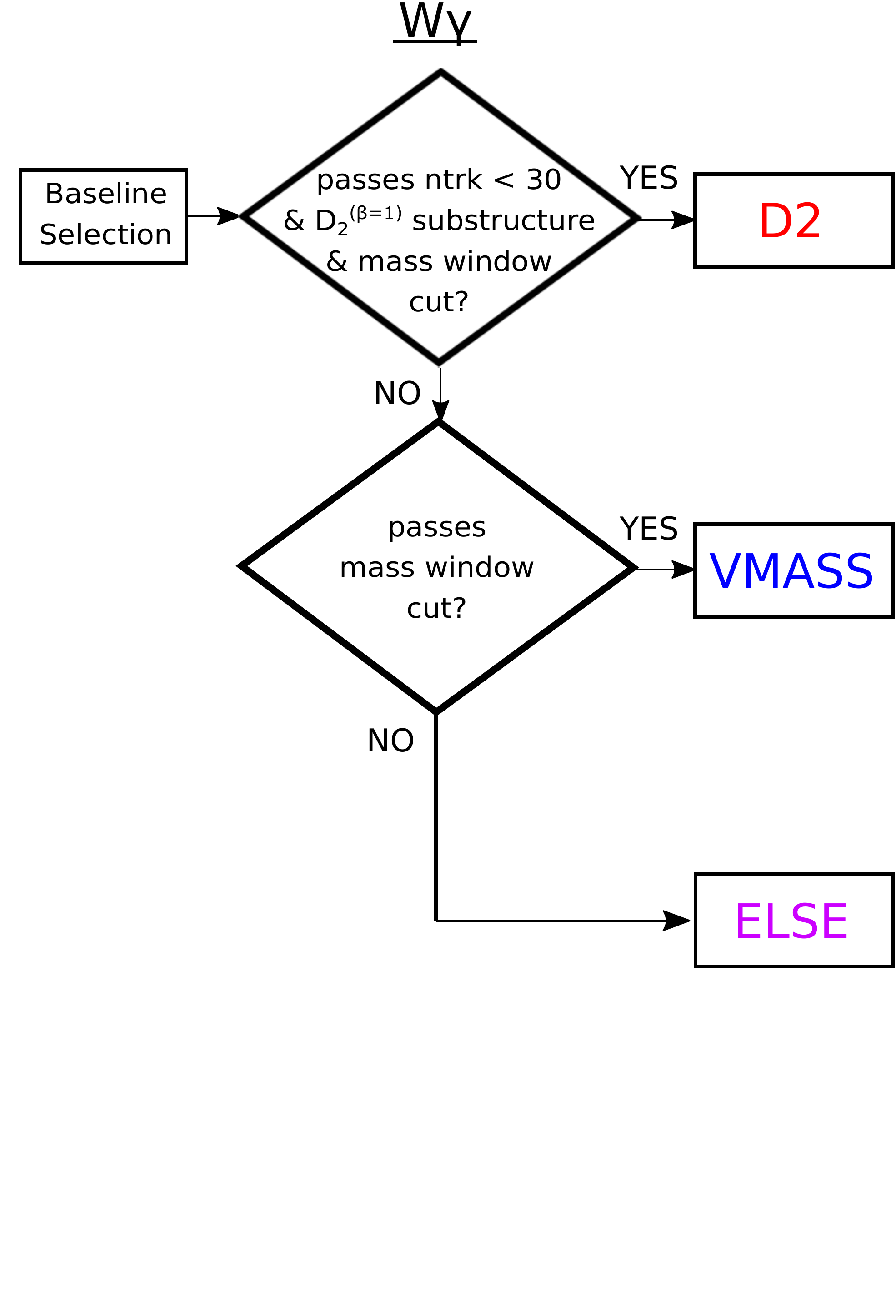}}
\end{center}
\caption{Flowcharts of the categorization of the events in (a) $Z\gamma$ and (b) $W\gamma$ searches. In the $H\gamma$ search only the BTAG selection is applied, analogous to that applied in the $Z\gamma$ search.}
\label{fig:cat_flow}
\end{figure}

\begin{figure}
\begin{center}
\subfigure[]{\includegraphics[width=0.48\columnwidth]{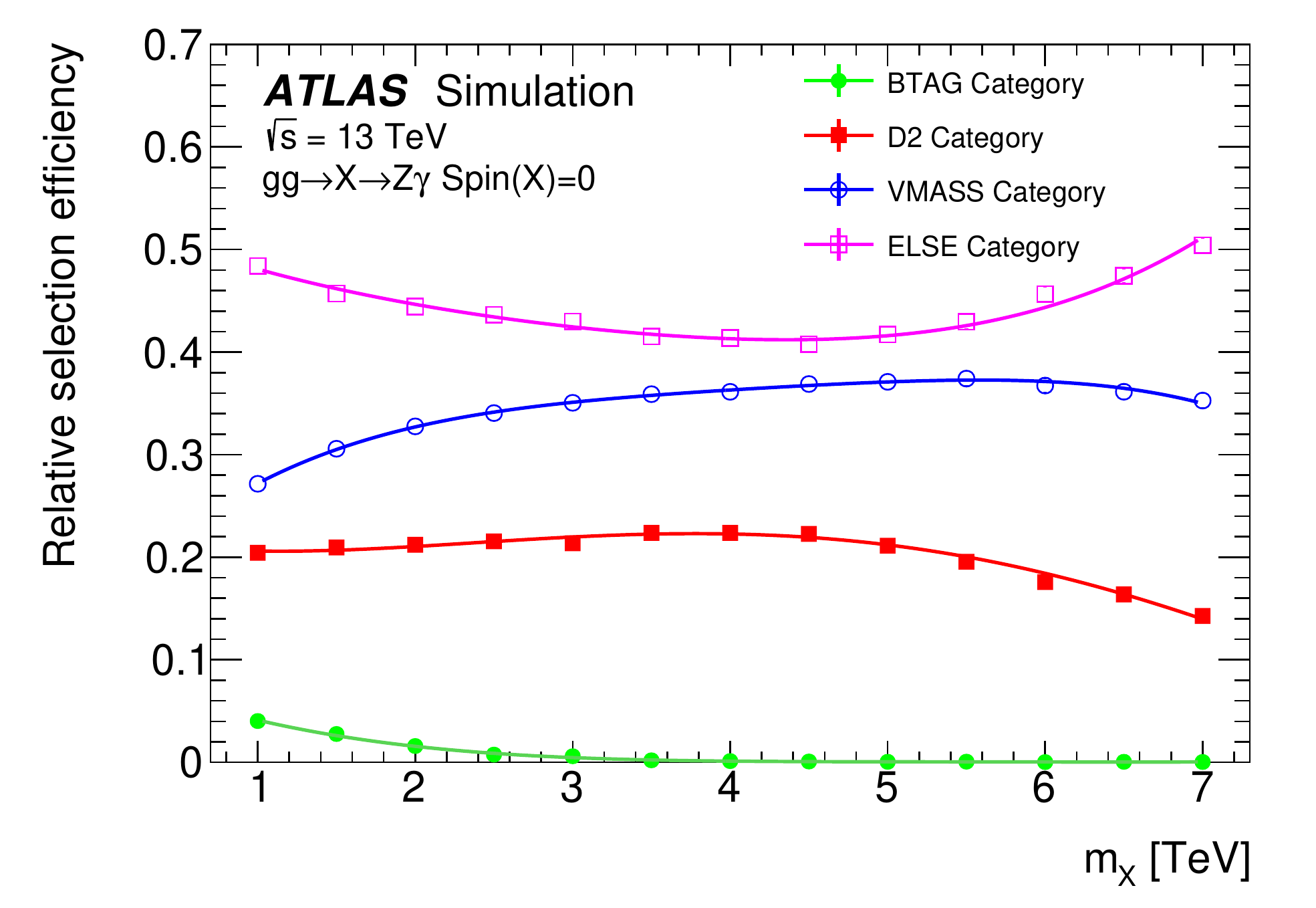}}
\subfigure[]{\includegraphics[width=0.48\columnwidth]{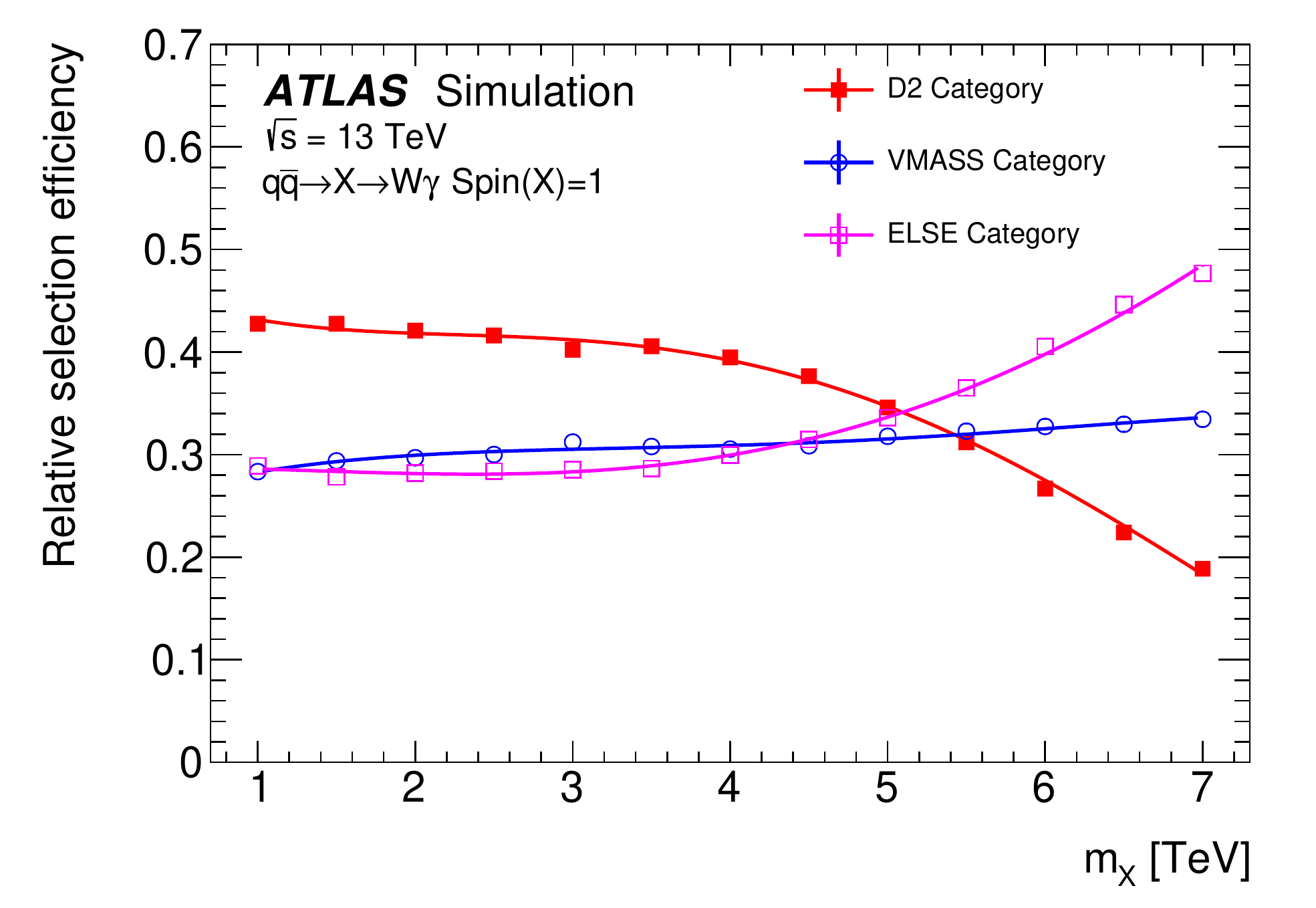}}
\subfigure[]{\includegraphics[width=0.48\columnwidth]{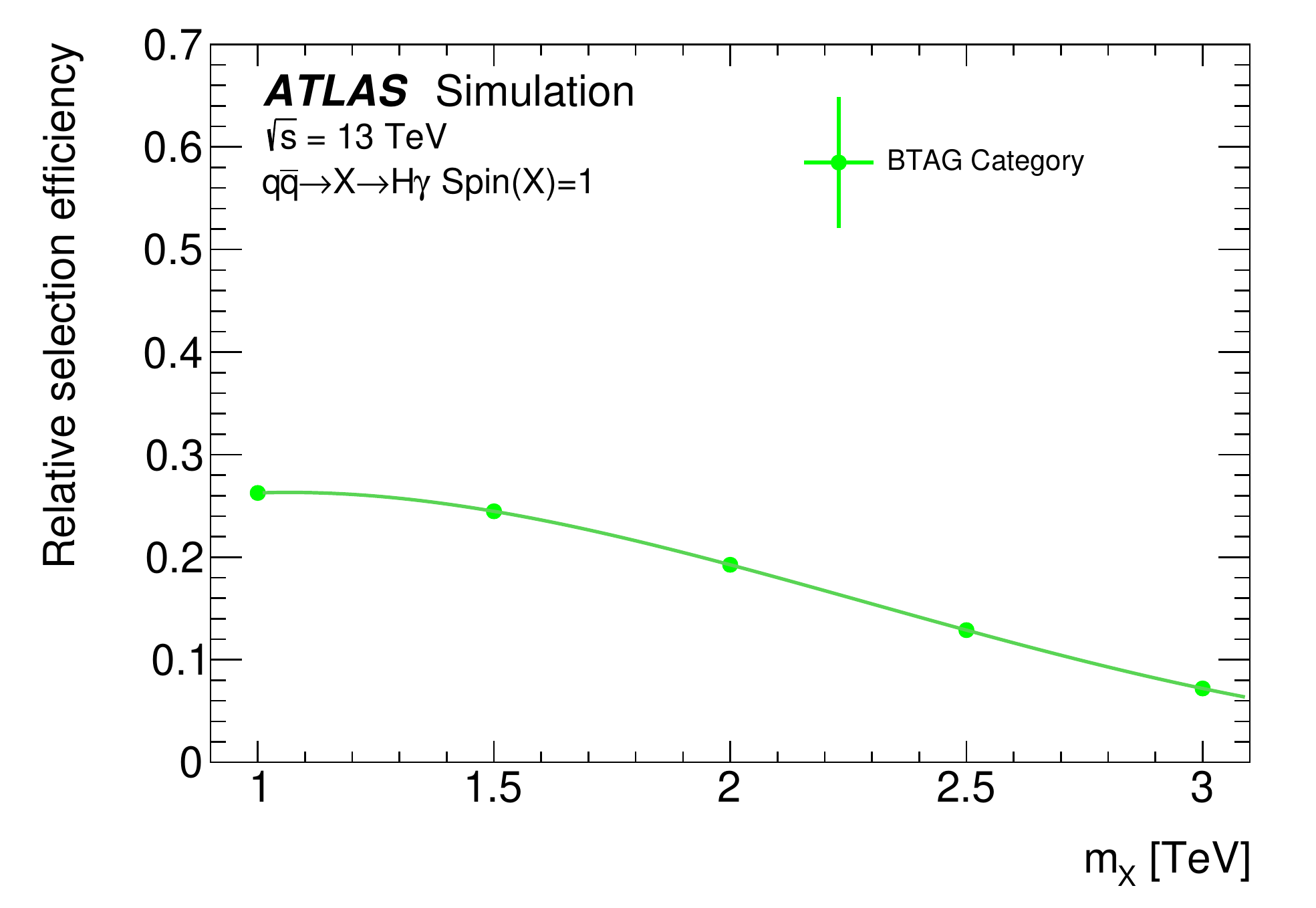}}
\end{center}
\caption{Efficiencies relative to the baseline selection for signal events to pass the category selection
  for (a) the $Z\gamma$ (in the spin 0 production hypothesis), (b) the
  $W\gamma$, and (c) the $H\gamma$ resonances as a function of the resonance mass
  $m_{X}$. The lines represent polynomial fits to the simulated data points. In the $Z\gamma$ search the categorization
  efficiencies have only a small dependence on the production and spin hypotheses.
  Statistical uncertainties are smaller than the marker size.}
\label{fig:cat_eff}
\end{figure}

\section{Signal and background models}
\label{sec:models}
The final discrimination between signal and background events
in the selected samples is achieved with a fit of a signal+background model to
the \mgj{} distribution
of the selected data events. The fit relies on the parameterization of signal and background \mgj{} distributions with a functional form.

\subsection{Signal model}
\label{ssec:signal_model}

The shape of the \mgj{} distribution is modeled by the sum of a Crystal Ball function~\cite{CrystalBall} representing the core populated by well-reconstructed events, and a Gaussian function modeling the tails populated by poorly reconstructed events as described by Eq.~(\ref{eqn:sig}):
\begin{align}
& S(\mgj{}) = f_{\mathcal{C}} \cdot \mathcal{C} (\mgj ;\mu,\sigma_{\mathcal{C}},\alpha_{\mathcal{C}},n_{\mathcal{C}}) + (1-f_{\mathcal{C}})\cdot \mathcal{G}(\mgj ;\mu,\sigma_{\mathcal{G}})\mathrm{,}
\label{eqn:sig}
\end{align}
    where $\mathcal{C}$ is the Crystal Ball function defined by Eq.~(\ref{eqn:CB}):
\begin{align}
\begin{split}
& \mathcal{C} (\mgj{};\mu,\sigma_{\mathcal{C}},\alpha_{\mathcal{C}},n_{\mathcal{C}}) \\
& \qquad = N_{\mathcal{C}} \cdot
	\begin{cases}
	\exp\left(- \frac{(\mgj{} - \mu)^2}{2\sigma_{\mathcal{C}}^2}\right) & \qquad\text{for } \frac{\mgj{} - \mu}{\sigma_{\mathcal{C}}} > - \alpha_{\mathcal{C}} \\
	\left( \frac{n_{\mathcal{C}}}{\alpha_{\mathcal{C}}} \right)^{n_{\mathcal{C}}} \exp\left(- \frac{\alpha_{\mathcal{C}}^2}{2}\right) \left( \frac{n_{\mathcal{C}}}{\alpha_{\mathcal{C}}} -\alpha_{\mathcal{C}} - \frac{\mgj{} - \mu}{\sigma_{\mathcal{C}}} \right) ^{-n_{\mathcal{C}}}     &  \qquad\text{for } \frac{\mgj{} - \mu}{\sigma_{\mathcal{C}}} \leq - \alpha_{\mathcal{C}}
	\end{cases}\mathrm{,}
\end{split}
\label{eqn:CB}
\end{align}
and $\mathcal{G}$ is a Gaussian function with mean $\mu$ and standard deviation $\sigma_{\mathcal{G}}$. The normalization factor $N_{\mathcal{C}}$ depends on the shape parameters of the Crystal Ball function.
The relative strength of the two distributions, $f_{\mathcal{C}}$, is a parameter of the signal description model.
Since the modeling of the tails of the mass peak is addressed by the Gaussian function, the parameter $n_{\mathcal{C}}$, controlling the shape of the tail of the Crystal Ball function, is fixed to 1.
Similarly, the parameter $\alpha_{\mathcal{C}}$, designating the onset of the power law tail of the Crystal Ball function is constrained to be between 0 and 4.
Additional free parameters $\sigma_{\mathcal{C}}$ and $\sigma_{\mathcal{G}}$ describe the widths of the Crystal Ball and the Gaussian distributions corresponding to the width of the core of the \mgj{} distribution and the width of the tails.

The model described above is fitted to the \mgj{} distribution of simulated events of each signal type considered, for $m_{\mathrm{X}}$ ranging from 850 \GeV{} to 7 \TeV{}, for each category.
To obtain a model varying continuously as a function of the resonance mass, the parameters are interpolated using polynomials of up to the third degree.
For all signal hypotheses, $\sigma_{\mathcal{C}}$ is about 20~\GeV{} for a 1~\TeV{} resonance, and grows linearly by 15~\GeV{} per 1~\TeV{} increase of the resonance mass.

\subsection{Background model}
\label{ssec:background_model}

Several SM processes contribute to the predicted event yield with
different proportions. In the BTAG category, the dominant SM process
is photon production in association with a $b$-flavored hadron,
whereas in the other categories photon production in association with
a light or $c$-flavored hadron dominates.
 The production of a photon in association with a $Z$ or $W$ boson, or the
pair production of top
 quarks also contribute to the total background. For events with $\mgj>1$ TeV the
 contribution in the BTAG category from $\gamma+Z$ production is
 approximately 32\% (13\%) for the $Z\gamma$ ($H\gamma$)
 search, while in the D2 category the contribution from $\gamma+W/Z$ production is approximately 15\% for both the
 $Z\gamma$ and $W\gamma$ searches. For other categories the
 contributions from SM $\gamma+W/Z$ production are below 5\%.

Samples of simulated events arising from the processes described
above are used to develop the functional modeling of the background and
to test the applicability of the functional form, number of
parameters, and the range of the fit. Multijet production, where one
jet is reconstructed as a photon, also contributes to the event samples.
The contribution of this type of events is estimated using a data-driven method~\cite{EXOT-2016-02} and shown to be about 10\% of the events passing the baseline selection and to not affect the \mgj  distribution.
Multijet production is therefore accounted for in the background model through the $\gamma$+jet production.

The family of functions from Ref.~\cite{EXOT-2015-02}, as described by Eq.~(\ref{back:eqn:background}), with up to five parameters, is used for the overall background model: 
\begin{equation}
  B(\mgj;\vecp) = (1-x)^{p_1}x^{p_2+p_3 \log(x)+p_4 \log^{2}(x)+p_5 \log^{3}(x)},
  \label{back:eqn:background}
\end{equation}
where $x$ is \mgj{} divided by the collision energy, and $\vecp\equiv(p_1,p_2,p_3,p_4,p_5)$ is the vector of shape parameters. 
In the VMASS and D2 categories, the fit spans the \mgj{} range of 800~\GeV{} to 7~\TeV{}; in the BTAG category, it spans the range from 800~\GeV{} to 3.2~\TeV; and in the ELSE category, it spans the range from 2.5~\TeV{} to 7~\TeV{}.


The number of parameters $p_i$ used in the model is chosen by testing the stability of the fit and the ability of the function to describe the \mgj{} background distributions over the range expected for the different event categories.
The MC-simulated backgrounds inside the signal region are used in these tests, and data outside the signal region are used to validate the functional form choice for background model.
The complementary data set selection follows the categorization procedure described in Section~\ref{sec:selection} with the exception that the photon candidates are required to satisfy 1.52 < $| \eta |$ < 2.37. The model also uses an F-test statistic~\cite{ftest} to decide on the minimum number of parameters required.
The number of model parameters $p_i$ varies from two to three depending on the event category.
The background modeling is stable while varying the relative fractions of the contributing SM processes.
Ensemble tests with pseudodata are used to validate the background model in regions of the \mgj{} distributions poorly populated by the data events.
The ensemble tests are performed separately in different categories for each of the $Z\gamma$, $W\gamma$, and $H\gamma$ searches.
Simulated event samples from SM processes are used to generate the pseudodata.
In addition ensemble tests are performed, where a signal process at the level of sensitivity of this search is also included in the simulated data.


\section{Systematic uncertainties}
\label{sec:systematics}
Uncertainties from systematic effects are due to the background estimation as well as the detector modeling, which affect the shape and normalization of the signal \mgj{} distributions.
These effects are estimated as relative uncertainties for the signal efficiency and the position and width of the signal peak for resonance masses ranging from 1~\TeV{} to 7~\TeV{} in 500~\GeV{} steps.
The impact of these effects on the \mgj{} distribution is evaluated using simulation.
Third-order polynomial interpolations are used to obtain relative variations of the signal efficiency and shape parameters due to each systematic uncertainty, for an arbitrary value of the resonance mass.


\paragraph{Uncertainty in background estimate:}

A systematic uncertainty associated with the background description arises from a potential bias in the estimated number of signal events due to the functional form chosen for the background paramaterization described by Eq.~(\ref{back:eqn:background}).
This effect, referred to as the spurious signal ($N_\text{spur}$), is studied using simulated background events (including \(\gamma+\)jet and \(W/Z +\) jet).
The bias is estimated by fitting the signal-plus-background model to simulated background \mgj{} distributions in each event category, for each signal hypothesis, with the sample's statistical uncertainties as expected in the data.
The absolute number of fitted signal events at a given $m_{X}$ hypothesis defines the number of spurious signal events $N_\text{spur}(m_{X})$.
The impact of uncertainties in the background composition has been studied by varying the fraction of the \(W/Z +\) jet and \(\gamma+\)$b/c$ jet backgrounds by 50\%, and found to be negligible in the spurious signal estimate.
The impact of the spurious signal uncertainty on the exclusion limits is discussed in Section~\ref{sec:results}.

\paragraph{Luminosity:} An integrated luminosity uncertainty of 2.1\% is derived, following a methodology similar to that detailed in Ref.~\cite{DAPR-2013-01},
from a calibration using beam separation scans performed in August 2015 and May 2016.

\paragraph{Jet energy scale and resolution:}
The uncertainties in the jet energy scale and resolution are estimated using $\gamma$+jet and dijet events in the data~\cite{ATLAS-CONF-2017-063}.

The impact of the systematic uncertainty in the jet energy scale is a shift of the peak position of the signal \mgj{} distribution by 1$-$3\%.
The signal mass resolution varies by 5\% in the low-mass region ($m_{X}<2.5\,\TeV{}$) and by 15\% in the high-mass region due to the systematic uncertainty in the jet energy resolution.
The impact on the signal efficiency from the jet energy uncertainty is 2$-$6\%.

\paragraph{Photon energy scale and resolution:}
The uncertainties in the photon energy scale and resolution are estimated using electron data samples with $Z\rightarrow ee$ events and high-purity photon samples with radiative $Z \to e e \gamma$ events~\cite{ATL-PHYS-PUB-2016-015}.
The impact of the systematic uncertainty in the photon energy scale is a shift of 0.5\% in the peak position of the signal \mgj{} distribution.
The signal mass resolution varies by 1\% in the low-mass region and by 3\% in the high-mass region due to the systematic uncertainty in the photon energy resolution.

\paragraph{Photon identification, isolation, and trigger efficiency:}
The uncertainties in the reconstruction, identification, isolation, and trigger efficiency for photons are determined from data samples of $Z \to \ell\ell \gamma$,  $Z \to ee$, and inclusive photon events, using the methods described in Ref.~\cite{PERF-2013-04}.
The impact on the signal efficiency from the photon identification, isolation, and trigger systematic uncertainties is found to be less than 1.5\%, 0.5\%, and 0.1\%, respectively.

\paragraph{Heavy-flavor jet identification:}
Uncertainties in the $b$-tagging efficiency for track-jets are derived from the uncertainties measured in dedicated heavy-flavor-enriched data samples, following the methodology described in Ref.~\cite{PERF-2012-04}.
The uncertainties are measured as a function of $b$-jet $\pT$ and range between 2\% and 8\% for track-jets with \pt{} $<$ 250 \GeV{}.
For track-jets with \pt{} $>$ 250 \GeV{} the uncertainty in the tagging efficiencies is extrapolated using simulation~\cite{PERF-2012-04}, and is approximately 9\% for track-jets with \pt{} $>$ 400 \GeV{}.
The impact of these uncertainties on the signal efficiency is 10$-$20\%.


\paragraph{Number of primary-vertex-originated tracks associated with the jet:}
The requirement on the number of tracks originating from the primary vertex and associated with the jet induces a 6\% systematic uncertainty in the signal efficiency,
as estimated from the comparison of data control samples and samples of simulated events~\cite{EXOT-2016-19}.

\paragraph{Scale and resolution of the jet mass and \decr{}:}

The uncertainties in the scale and resolution of the jet mass and \decr{} are evaluated by comparing the ratio of the calorimeter-based to the track-based measurements in dijet data and simulation~\cite{ATL-PHYS-PUB-2015-033, PERF-2012-02}.
The jet mass resolution uncertainty affects the signal efficiency by 3$-$6\%.
The impact of \decr{} scale and resolution uncertainties on the signal efficiency and the core width~$\sigma_{\mathcal{C}}$ of the signal peak is found to be less than 1\%.

\paragraph{Pileup modeling:}
The pileup weighting of the simulated signal events, described in Section~\ref{sec:data},
is varied to cover the uncertainty in the ratio of the predicted and measured inelastic $pp$ cross sections~\cite{STDM-2015-05}. 
The pileup uncertainty affects the signal efficiency by 1$-$2\%.

\paragraph{PDF choice:}
The uncertainty due to the PDF modeling is evaluated by comparing the signal acceptances for alternative PDF sets to that for the nominal set.
The total uncertainty in the acceptance is derived as the standard deviation of the eigen-variations according to the method described in Ref.~\cite{Buckley:2014ana}.
The uncertainties in acceptances for signal processes produced via $q\bar{q}$ annihilation vary from 5\% to 2\% with increasing
resonance mass, while for signal processes produced via gluon--gluon fusion the uncertainties vary from 12\% to 2\%.

\paragraph{Parton shower:}
The uncertainty due to the parton shower modeling is evaluated by comparing the signal acceptances for alternative parton shower models to the acceptance for the nominal model.
The alternative parton shower models are defined as the eigen-variations of the $\textsc{Pythia}$ A14 tune~\cite{ATL-PHYS-PUB-2014-021}.
The total uncertainty is derived as the sum in quadrature of all the eigen-variation effects and affects the acceptance by about $2\%$.

The effects of systematic shifts on signal normalization, position of the signal peak,
and the core width $\sigma_{\mathcal{C}}$ of the signal peak are summarized in Table~\ref{tab:sys} for the $Z\gamma$, $W\gamma$, and $H\gamma$ searches.
The effects on the signal acceptances from theoretical uncertainties are also given.

\begin{table}[htbp]
\centering

\begin{tabularx}{0.98\textwidth}{cccccccc}
\toprule
\midrule
    & \multicolumn{1}{c}{Impact on normalization and efficiency [\%]} \\
\hline
Luminosity                   & 2.1  \\
Jet energy scale     &  2--6  \\
Photon identification and isolation         & 0.5--1.5 \\

Flavor tagging                  &   10--20  \\ 
$n_{\mathrm{trk}}$ associated with the jet   & 6 \\
Jet mass resolution          & 3--6   \\
\decr{} scale and resolution                & $<1$ \\
Pileup modeling    & 1--2 \\
\hline
 & \multicolumn{1}{c}{Impact on signal peak position} [\%] \\
\hline

Jet energy and mass scale     &  1--3  \\
Photon energy scale  & $<0.5$    \\
\hline
& \multicolumn{1}{c}{Impact on signal peak resolution} [\%]\\
\hline
Jet energy resolution         & 5 ($m_X<$ 2.5~\TeV{})--15 ($m_X>$ 2.5~\TeV{}) \\
Photon energy resolution & 1--3 \\
\hline
& \multicolumn{1}{c}{Impact on acceptance} [\%]\\
\hline
PDF  & 2--12 \\
Parton shower & 2 \\
\bottomrule
\end{tabularx}

\caption{Effect of systematic uncertainties from various sources on signal normalization and efficiency, position of the signal peak, and the core width $\sigma_{\mathcal{C}}$ of the signal peak.
The last two rows show the theoretical uncertainty effects on the signal acceptances.}
\label{tab:sys}
\end{table}

\section{Statistical procedure}
\label{sec:statistics}

The data are scrutinized with statistical methods to quantify the
presence of a hypothetical resonance and to set a limit on its
production.
In both cases an unbinned extended maximum likelihood estimator is used to model
the data. The parameter of interest is \sbhad{}, the
signal cross section times the branching fraction of the resonance
decaying to $(Z\mathrm{/}W\mathrm{/}H)\gamma$ with subsequent hadronic
($b\bar{b}$) decays of $Z/W$ (Higgs) bosons. The impact of systematic
uncertainties on the signal is modeled with a vector of nuisance
parameters, \vecth{}, where each component, \thetak{}, is constrained
with corresponding Gaussian probability density functions
$\mathcal{G}_k(\thetak)$. The likelihood model, $ \mathcal{L} $, for
the sample of data events is described by
Eq.~(\ref{eqn:likelihood}): 
  \begin{equation}
    \begin{aligned}
\mathcal{L}(\sbhad , m_{X}) =& \prod_{j} \left[ \frac{\mathrm{e}^{-N_j(\sbhad, \vecth)}N_j(\sbhad, \vecth)^{n_j}}{n_{j}!} \prod_{\ell} \left(f_{\mathrm{tot},j}(\mgjl; \sbhad, \vecp, \vecth, m_{X})\right)\right]
  &\prod_{k} \mathcal{G}_k(\thetak),
 \end{aligned}
 \label{eqn:likelihood}
 \end{equation} 
where $j$ represents the event category, $n_{j}$ is the observed number of events in that category, $\ell$ is the event index, and $N_j(\sbhad, \vecth)$ is the expected total event yield in category $j$.
The total probability density function $f_{\mathrm{tot},j}$ in category $j$ depends on the photon-jet invariant mass \mgjl{} and is a function of the parameter of interest \sbhad{}, the parameters of the background modeling function \vecp, the nuisance parameters \thetak{}, as well as the mass $m_{X}$ of the hypothetical resonance. The functional form of $f_{\mathrm{tot},j}$ is given in Eq.~(\ref{eqn:ftot}):
\begin{equation}
    \begin{aligned}
f_{\mathrm{tot},j}(\mgjl; \sbhad, \vecp, \vecth, m_{X})=\frac{1}{N_j(\sbhad, \vecth, \vecp)}  \left[ \left( N_{\mathrm{sig}, j}(\sbhad, \vecth,m_{X}) + N_{\mathrm{spur}, j}(\theta_{\mathrm{spur}, j},m_{X}) \right) \times \right. \\
\left. S_j(\mgjl; \sbhad, \vecth) + N_{\mathrm{bkg}, j} B_j(\mgjl; \vecp_j)\right],
\end{aligned}
\label{eqn:ftot}
\end{equation}
where $S_j$ and $B_j$ are the signal and background probability density functions in category $j$, described in Section~\ref{sec:models}.
The parameter $\theta_{\mathrm{spur}, j}$ is an element of the \vecth vector corresponding to the spurious signal nuisance parameter.
The expected yield of signal events $N_{\mathrm{sig}, j}$ is given by the product of \sbhad{}, integrated luminosity, acceptance and efficiency for a given category $j$.
The expected number of background events $N_{\mathrm{bkg}, j}$ is a parameter of the fit. The total expected event yield in a given category, $N_j$, is the sum of the expected signal, background, and spurious signal event yields.

The $p$-values are computed to examine the compatibility of the data and the background-only hypothesis.
First, the local $p$-value is calculated for the particular value of $m_{X}$ under consideration.
The local $p$-value is defined as the probability of the background to produce a signal-like excess whose estimated \sbhad{} is larger than that found in the fit to the data in all categories simultaneously.
This procedure utilizes the ratio of the likelihood value where the most likely value of the parameter of interest \sbhad{} is found, to the likelihood value where no signal is allowed (\sbhad{} = 0)~\cite{stat}.
The global $p$-value is defined as the probability of finding, at any value of  $m_{X}$, a signal-like fluctuation more significant than the most significant excess found in the data, in all categories combined. It is calculated approximately by discounting the local $p$-value by the effective number of search trials possible within the \mgj{} range examined.

The modified frequentist (CL$_\text{s}$) method~\cite{cls,cls_junk} is used to set
upper limits on the signal \sbhad{} at 95\%
confidence level (CL).
 To obtain limits on  $\sigma B$, which is the signal cross section times the branching fraction of resonance decays to $(Z\mathrm{/}W\mathrm{/}H)\gamma$ with all allowed decays of $Z$, $W$,
and Higgs bosons, the resulting $\sigma B_{\mathrm{had}}$ is divided by the measured hadronic branching fraction of the $Z$ (69.91\%~\cite{Patrignani:2016xqp}) or $W$ (67.41\%~\cite{Patrignani:2016xqp}) bosons or the theoretically calculated branching fraction of the Higgs boson to $b\bar{b}$ (58.24\%~\cite{deFlorian:2016spz}).

 Closed-form asymptotic formulae~\cite{stat} are used to calculate the limits. The $p$-value calculation and $\sigma B$ limit calculations are performed for $m_{\mathrm{X}}$ in range of 1 \TeV{} to 6.8~\TeV{} in steps of 20~\GeV{}. The step size is chosen to be much smaller than the experimental \mgj{} resolution and the interval spans the interpolation range of the shape parameters. The VMASS and D2 categories are used in the entire range, while the BTAG category is used only for 1~\TeV{} < $m_{\mathrm{X}}$ < 3~\TeV{}, and the ELSE category is used only for 3~\TeV{} < $m_{\mathrm{X}}$ < 6.8~\TeV{}. The fit interval described in Section~\ref{ssec:background_model} is selected such that a peak in the \mgj{} distribution, resulting from resonance with $m_{\mathrm{X}}$ considered, is fully contained.

Due to the small number of events for large $m_{X}$ values, the results are checked with ensemble tests.
The limits derived with asymptotic formulae agree well with those obtained with ensemble tests for small $m_{X}$ values,
and are underestimated by as much as 30\% for large $m_{X}$ values.


\section{Results}
\label{sec:results}

The data event yields in the baseline selection as well as in all search channels and in the categories within each channel are shown in Table~\ref{tab:yields}.

The observed \mgj{} distributions as well as the background distributions obtained from a global background-only fit with various hypothetical signal curves overlaid are shown in Figures.~\ref{fig:dataZgamma},~\ref{fig:dataWgamma}, and~\ref{fig:dataHgamma} for the $Z\gamma$, $W\gamma$, and $H\gamma$ searches, respectively. The event with the highest \mgj{}, at a value of 6.3~\TeV{}, has a jet of mass 63~\GeV{} and it is found in the ELSE (VMASS) category of the $Z\gamma$ ($W\gamma$) search.

\begin{table}[h]
\centering
\begin{tabular}{l ccccc}
\toprule
\multirow{2}{*}[-0.7ex]{Selection} & \multicolumn{5}{c}{Event yield in each category ($\mgj>$ 1 TeV)} \\
  \cmidrule(l{2pt}r{2pt}){2-6}
& Baseline & BTAG & D2 & VMASS & ELSE \\
\midrule
$Z\gamma$ search & 60,237 & 25 & 784 & 5,569 & 53,859 \\
\midrule
$W\gamma$ search & 60,237 & $\cdots$ & 661 & 5,216 & 54,360 \\
\midrule
$H\gamma$ search & 60,237 & 59 & $\cdots$ & $\cdots$ & $\cdots$ \\
\bottomrule
\end{tabular}
\caption{Event yields in the baseline selection and in the $Z\gamma$, $W\gamma$, and the $H\gamma$ searches, in the categories used in those searches. Only events with $\mgj>$ 1 TeV are considered.}
\label{tab:yields}
\end{table}

\begin{figure}[!h]
  \begin{center}
    \subfigure[]{\includegraphics[width=0.48\columnwidth]{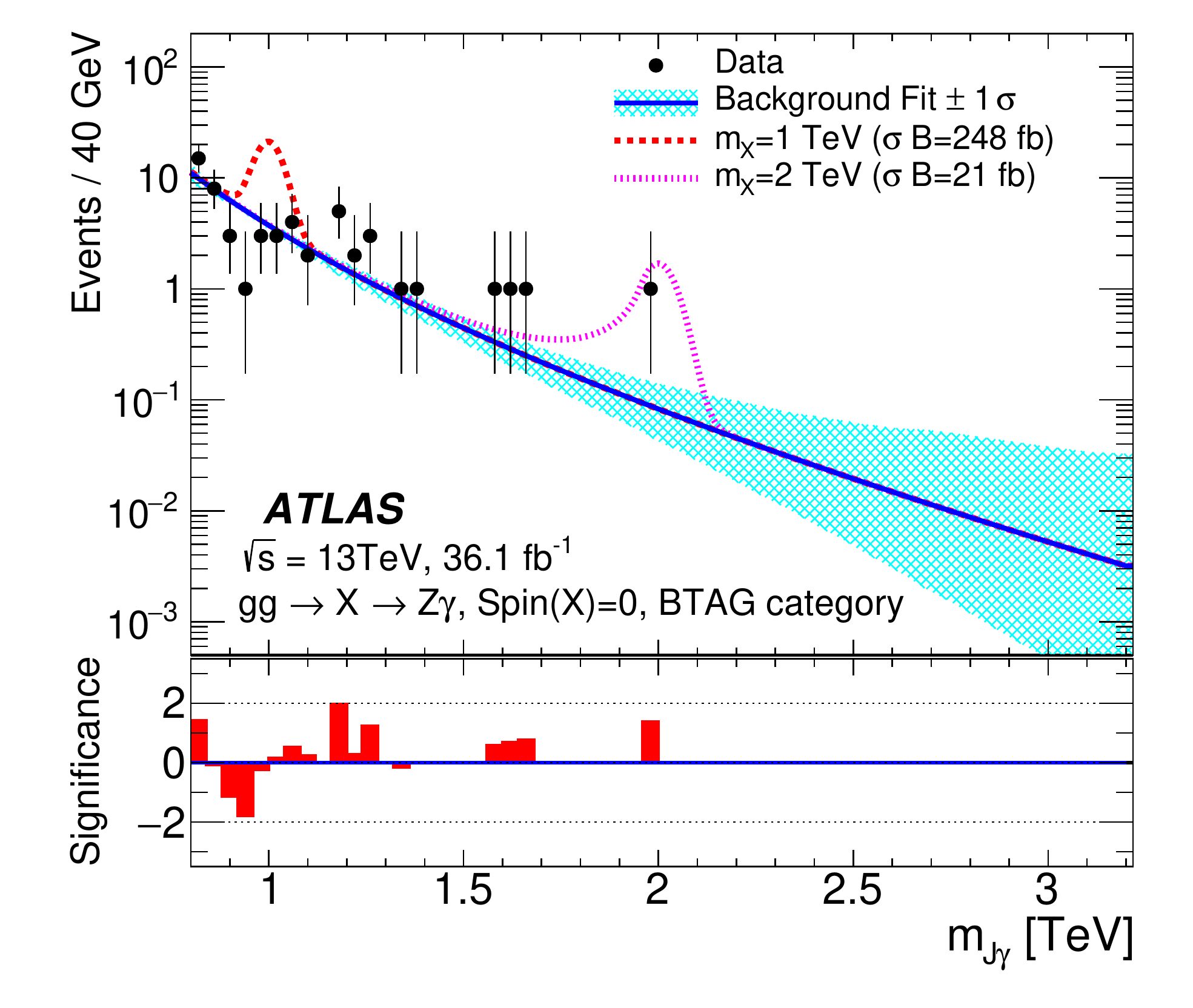}}
    \subfigure[]{\includegraphics[width=0.48\columnwidth]{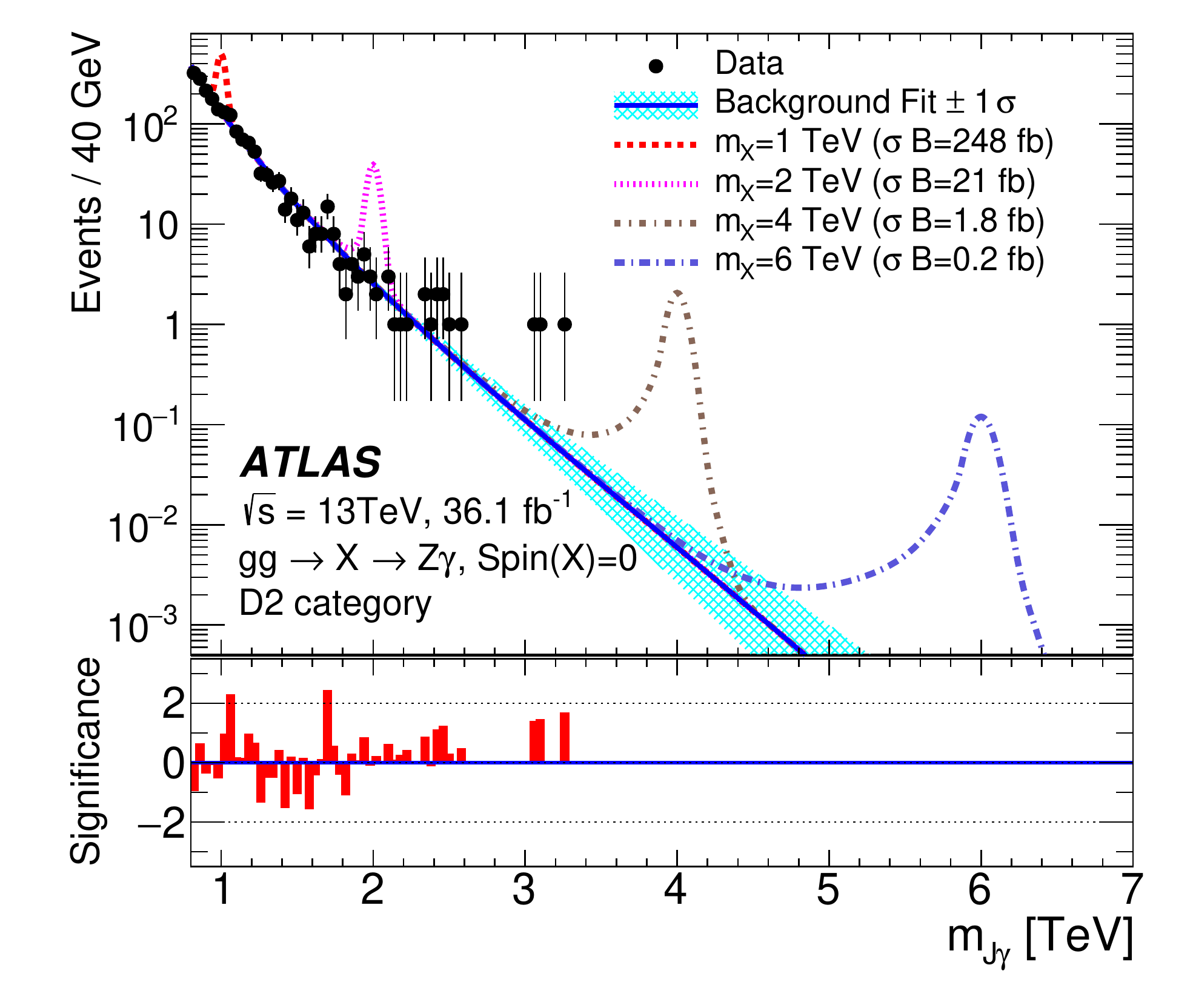}}
    \subfigure[]{\includegraphics[width=0.48\columnwidth]{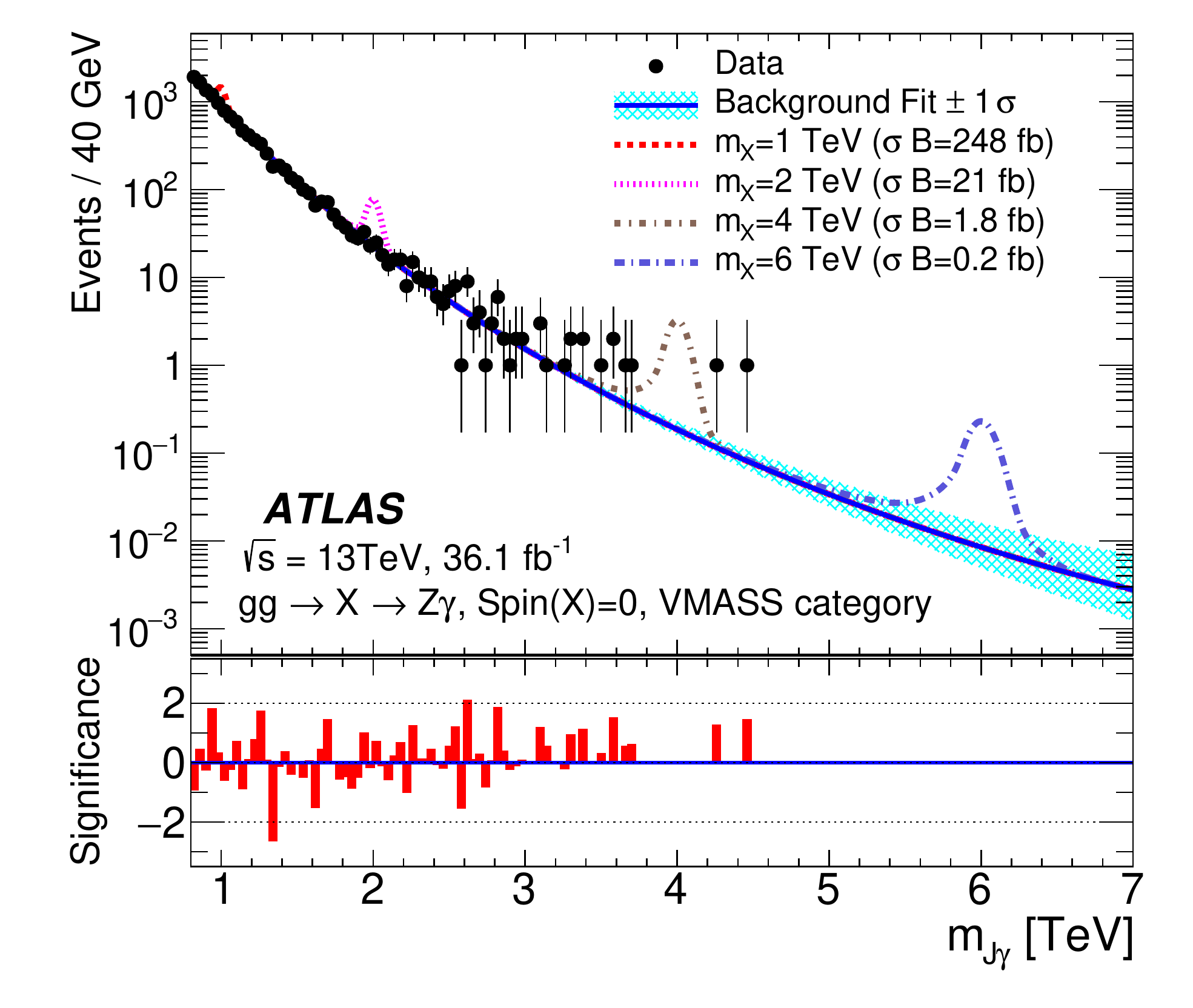}}
    \subfigure[]{\includegraphics[width=0.48\columnwidth]{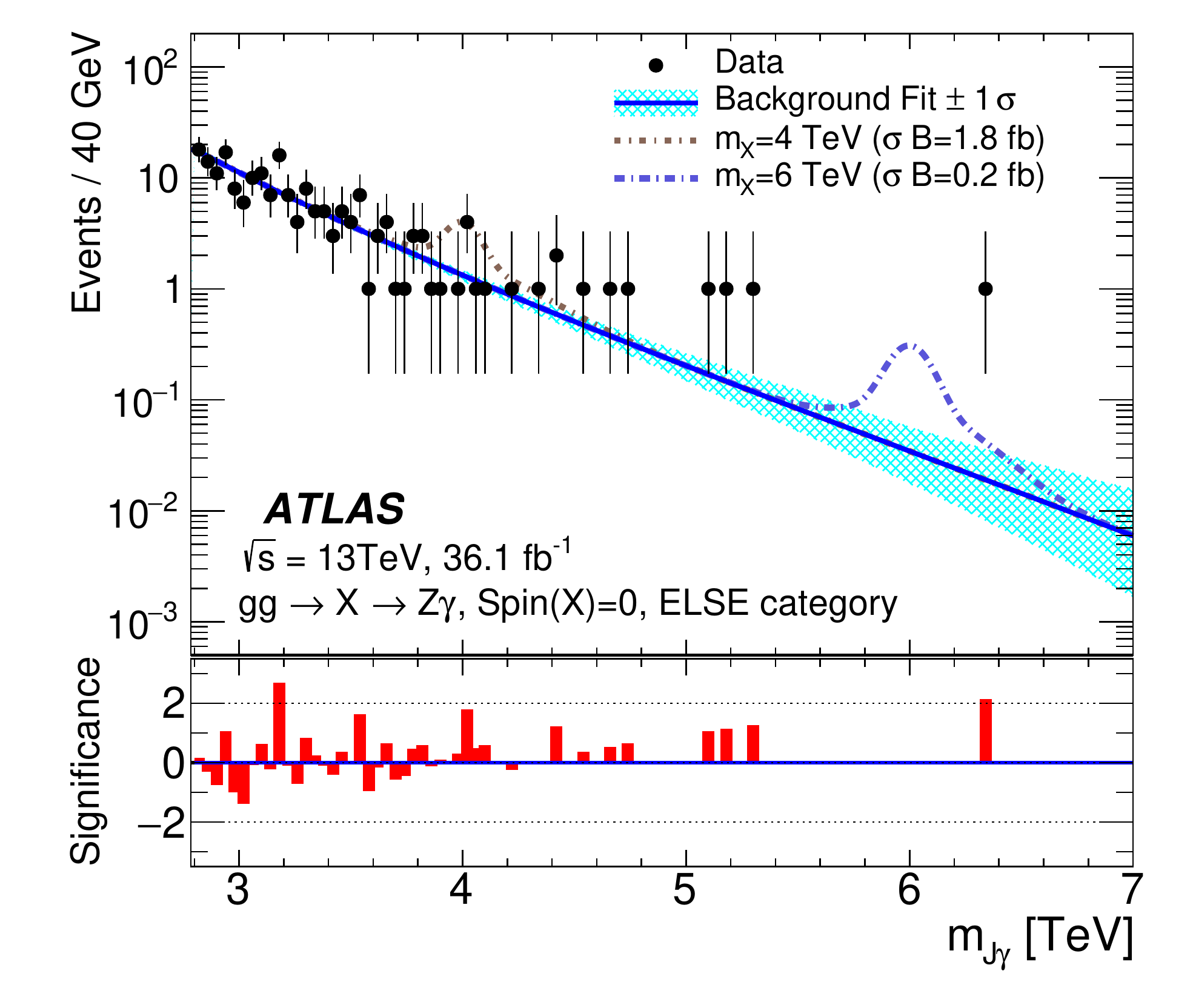}}
    \caption{Distributions of the reconstructed mass \mgj{} in the $Z\gamma$ search (a) BTAG, (b) D2, (c) VMASS, and (d) ELSE categories.
        The models obtained in the background-only fits are shown by the solid lines. 
        Hypothetical signal distributions with $\sigma B$ at the level
        excluded multiplied by factors of 20~(for 1~\TeV{}), 10~(for 2~\TeV{}),
        5~(for 4~\TeV{}), and 1~(for 6~\TeV{}) for the given signal
        model and resonance mass are overlaid. The $\sigma B$ lines are calculated with the scale factors applied. The bottom panels give
  the significance for each bin. The significance calculation assumes that the background estimate in a given bin is Poisson distributed and follows the recommendation of Ref.~\cite{binned_significance}.
  The impact on the background fit of the statistical uncertainties in parameters $p_i$ is shown as a light band around the solid line.
  This effect is incorporated into the significance calculation.
  }
    \label{fig:dataZgamma}
  \end{center}
\end{figure}

\begin{figure}[!h]
  \begin{center}
\subfigure[]{\includegraphics[width=0.48\columnwidth]{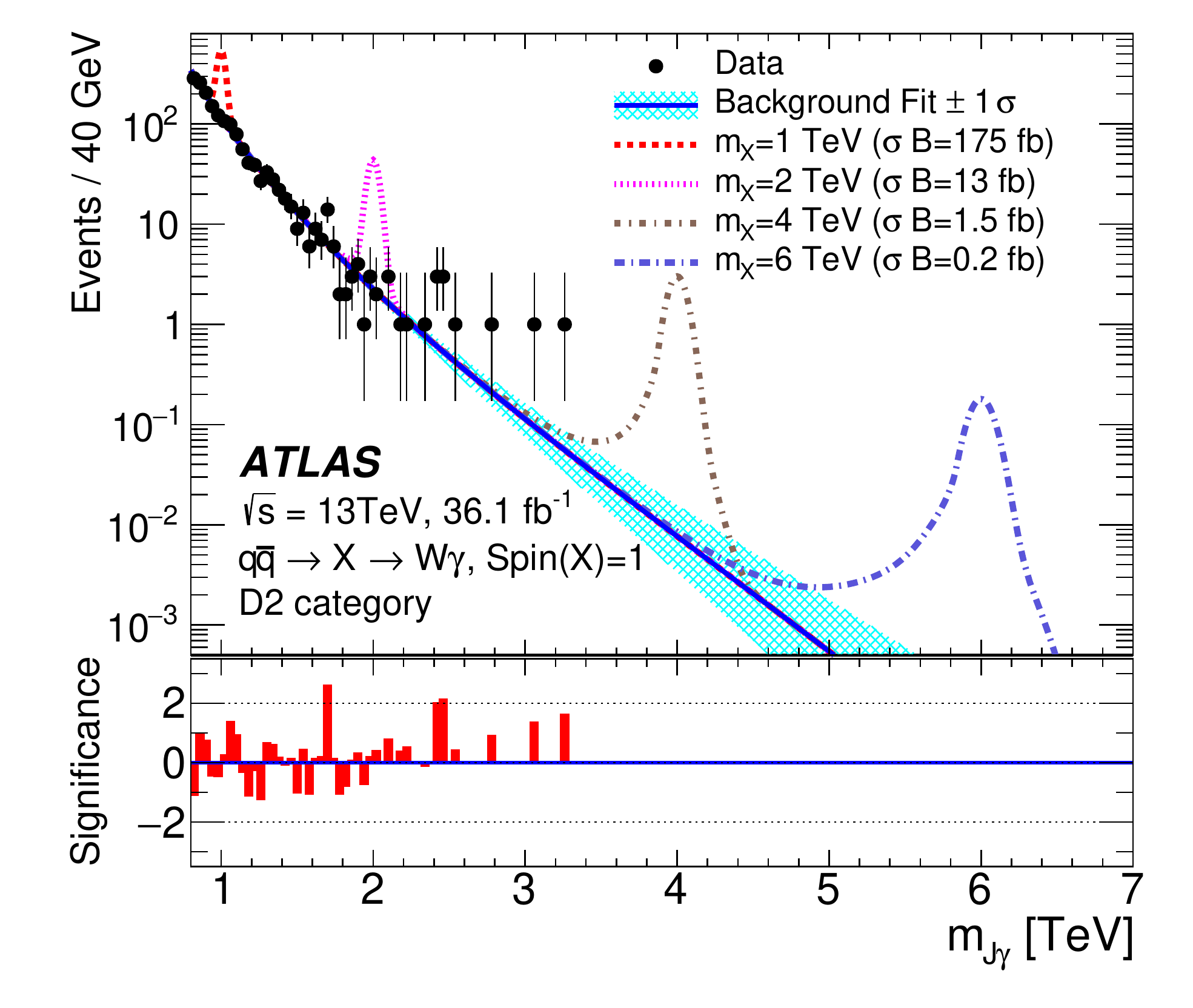}}
\subfigure[]{\includegraphics[width=0.48\columnwidth]{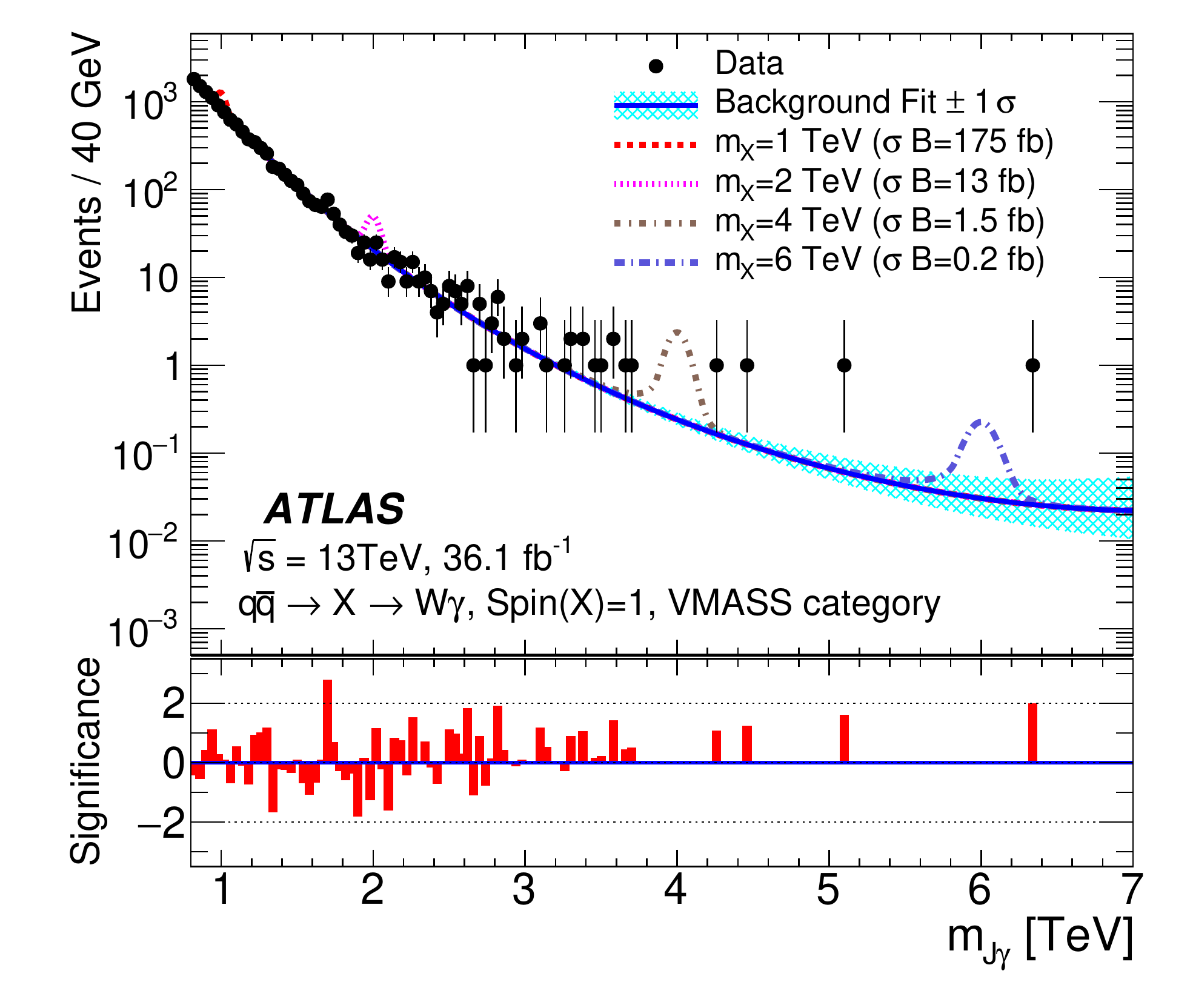}}
\subfigure[]{\includegraphics[width=0.48\columnwidth]{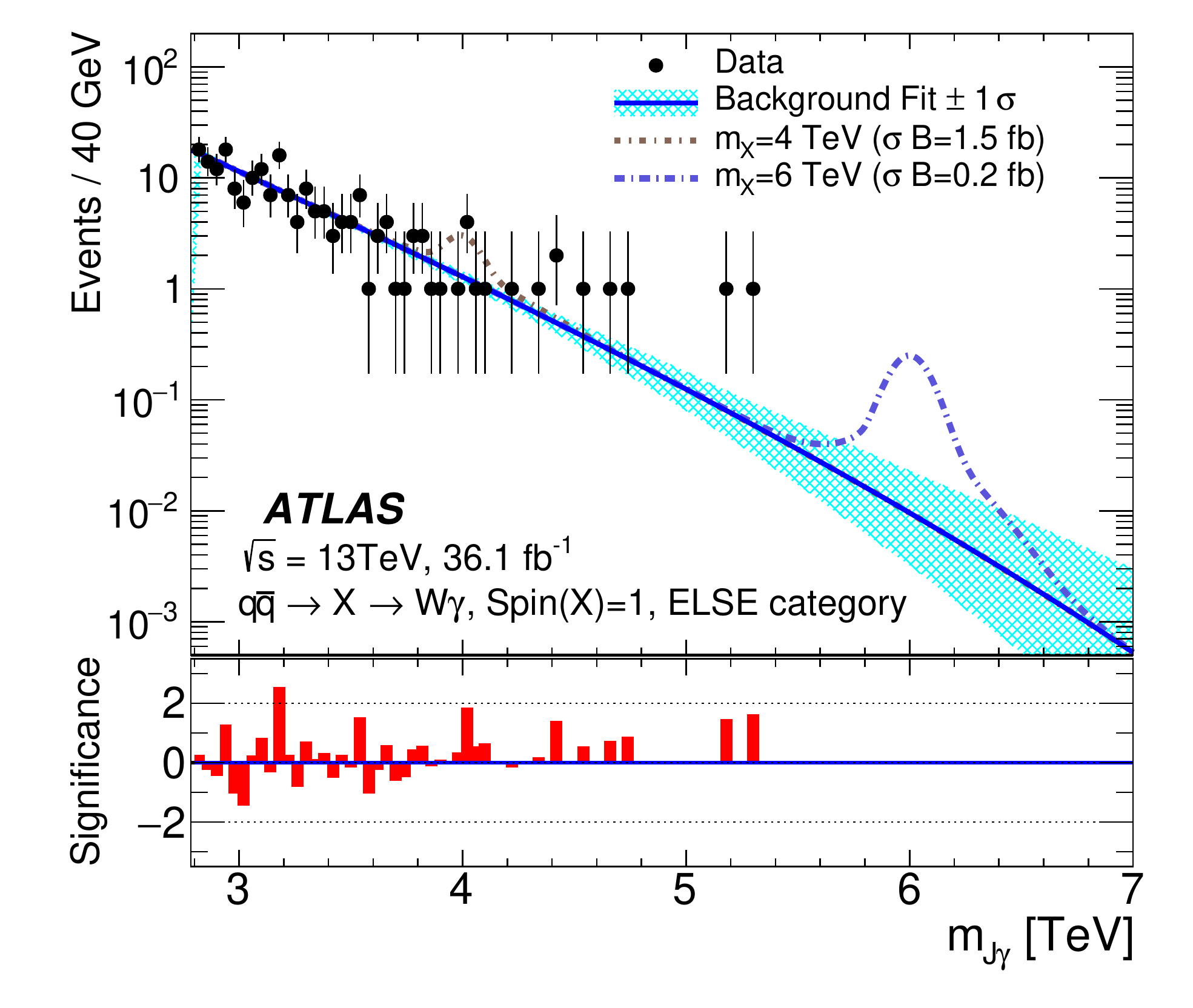}}
      \caption{Distributions of the reconstructed mass \mgj{} in the $W\gamma$ (a) D2, (b) VMASS, and (c) ELSE categories.
        The models obtained in the background-only fits are shown by the solid lines. 
        Hypothetical signal distributions with $\sigma B$ at the level
        excluded multiplied by factors of 20~(for 1~\TeV{}), 10~(for 2~\TeV{}),
        5~(for 4~\TeV{}), and 1~(for 6~\TeV{}) for the given signal
        model and resonance mass are overlaid. The $\sigma B$ lines are calculated with the scale factors applied. The bottom panels give
        the significance for each bin. The significance calculation assumes that the background estimate in a given bin is Poisson distributed. The calculation follows the recommendation of Ref.~\cite{binned_significance}. The impact on the background fit of the statistical uncertainties in parameters $p_i$ is shown as a light band around the solid line. This effect is incorporated into the significance calculation.
}
    \label{fig:dataWgamma}
  \end{center}
\end{figure}

\begin{figure}[!h]
  \begin{center}
    \includegraphics[width=0.48\columnwidth]{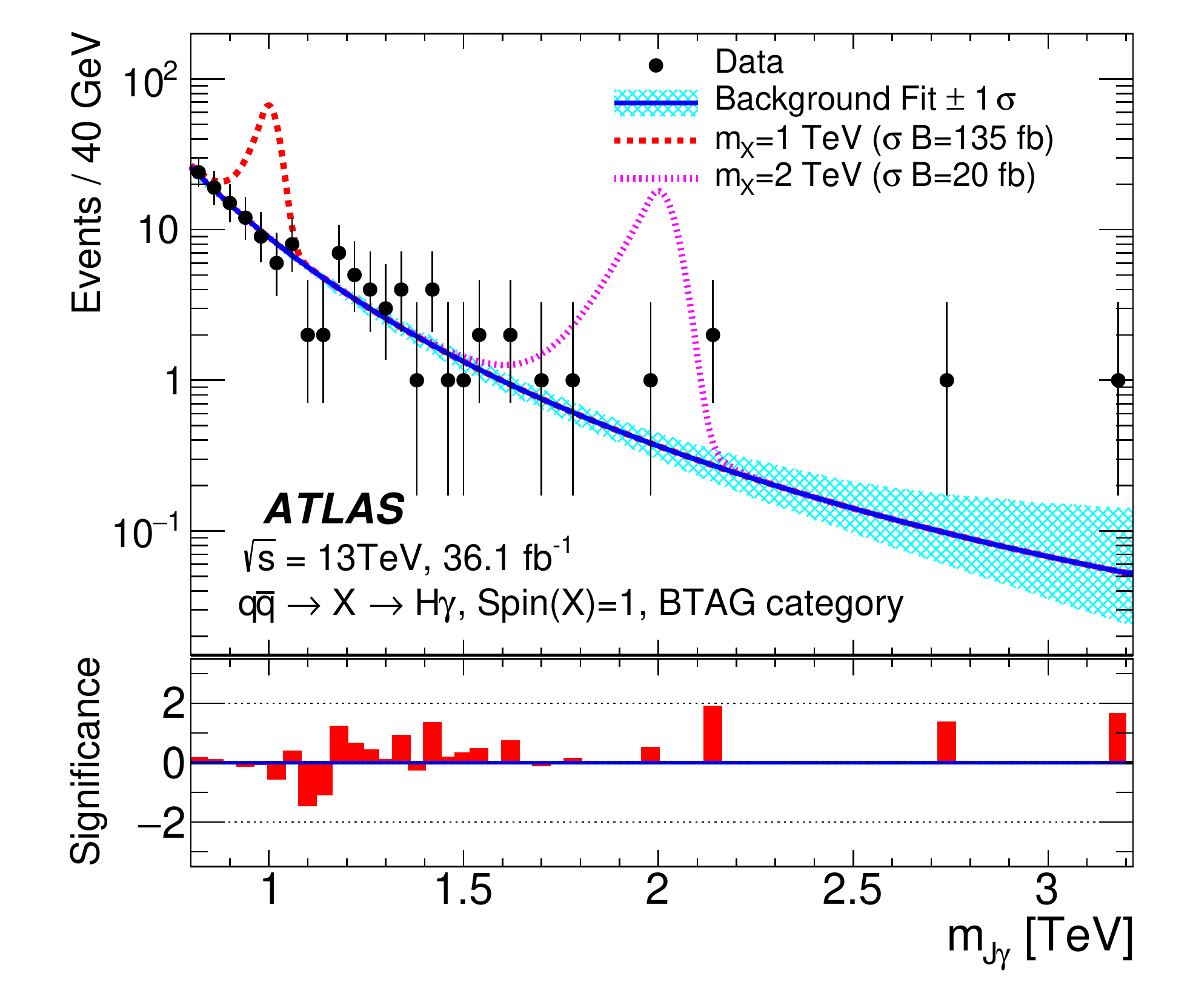}
      \caption{Distribution of the reconstructed mass \mgj{} in the $H\gamma$ search BTAG category.
        The models obtained in the background-only fits are shown by the solid lines. 
        Hypothetical signal distributions with $\sigma B$ at the level
        excluded multiplied by factors of 20~(for 1~\TeV{}) and 10~(for 2~\TeV{}) for
        the given signal model and resonance mass are overlaid. The $\sigma B$ lines are calculated with the scale factors applied. The
        bottom panel gives the significance for each bin. The significance calculation assumes that the background estimate in a given bin is Poisson distributed. The calculation follows the recommendation of Ref.~\cite{binned_significance}. The impact on the background fit of the statistical uncertainties in parameters $p_i$ is shown as a light band around the solid line. This effect is incorporated into the significance calculation.
    }
    \label{fig:dataHgamma}
  \end{center}
\end{figure}




The smallest local $p$-value, corresponding to a significance of $2.7\,\sigma$, is found in the $W\gamma$ search at \newline
\mgj{} $= 2.5$~\TeV{} utilizing data from all categories simultaneously. This local $p$-value corresponds to
a global significance of less than $1\sigma$. No significant excess is observed
in any of the categories and analysis channels. Limits are placed on specific models.

The $\sigma B$ limits on $Z\gamma$ production, evaluated for resonance masses between 1.0 and 6.8~\TeV{}, are shown in Figure~\ref{fig:limitsZ}. Spin and production hypotheses comprise a spin-0 resonance, produced by gluon--gluon fusion and a spin-2 resonance, produced by gluon--gluon fusion and $q\bar{q}$ annihilation. The $\sigma B$ limit on $W\gamma$ resonances evaluated for $m_{X}$ between 1 and 6.8~\TeV{} are shown in Figure~\ref{fig:limitsW}. This is the first evaluation of such a limit utilizing hadronic $W$ boson decays. The $Z\gamma$ and $W\gamma$ limits decrease from approximately 10 fb for a resonance mass of 1~\TeV{} to 0.1 fb for a resonance mass of 6.8~\TeV{}.  The $H\gamma$ search presented here is the first search for a heavy resonance with this decay mode. The $\sigma B$ limits on the resonances decaying to a $H\gamma$ final state are shown in Figure~\ref{fig:limitsH}. The limit is evaluated for resonance masses between 1 and 3~\TeV{} and varies between 10 fb and 4 fb depending on $m_{X}$. The limit weakens for $m_{X}$ > 2~\TeV{} due to a decrease in signal efficiency as shown in Figure~\ref{fig:cat_eff}, stemming from the decrease in the $b$-tagging efficiency for high-momentum jets.

The sensitivity of the resonance search and the strength of the resonance production cross section limit are primarily determined by the available data sample size.
Among all the systematic uncertainties, the spurious-signal uncertainty on the
background estimation has the largest impact on the limit, in particular in the
low-mass region,
weakening the limit by up to 20\% (1\%) at $m_{X}=1$~\TeV{} (6.8~\TeV{}). Another important systematic uncertainty is that in the heavy-flavor jet identification efficiency. It weakens the limit by up to 13\% (20\%) at $m_{X}=1$~\TeV{} (3~\TeV{}) in the $H\gamma$ analysis, while it has little impact on the limits in the $Z\gamma$ analysis since the BTAG category is just one of the four categories in this analysis.

\begin{figure}[!h]
  \begin{center}
    \subfigure[]{\includegraphics[width=0.48\columnwidth]{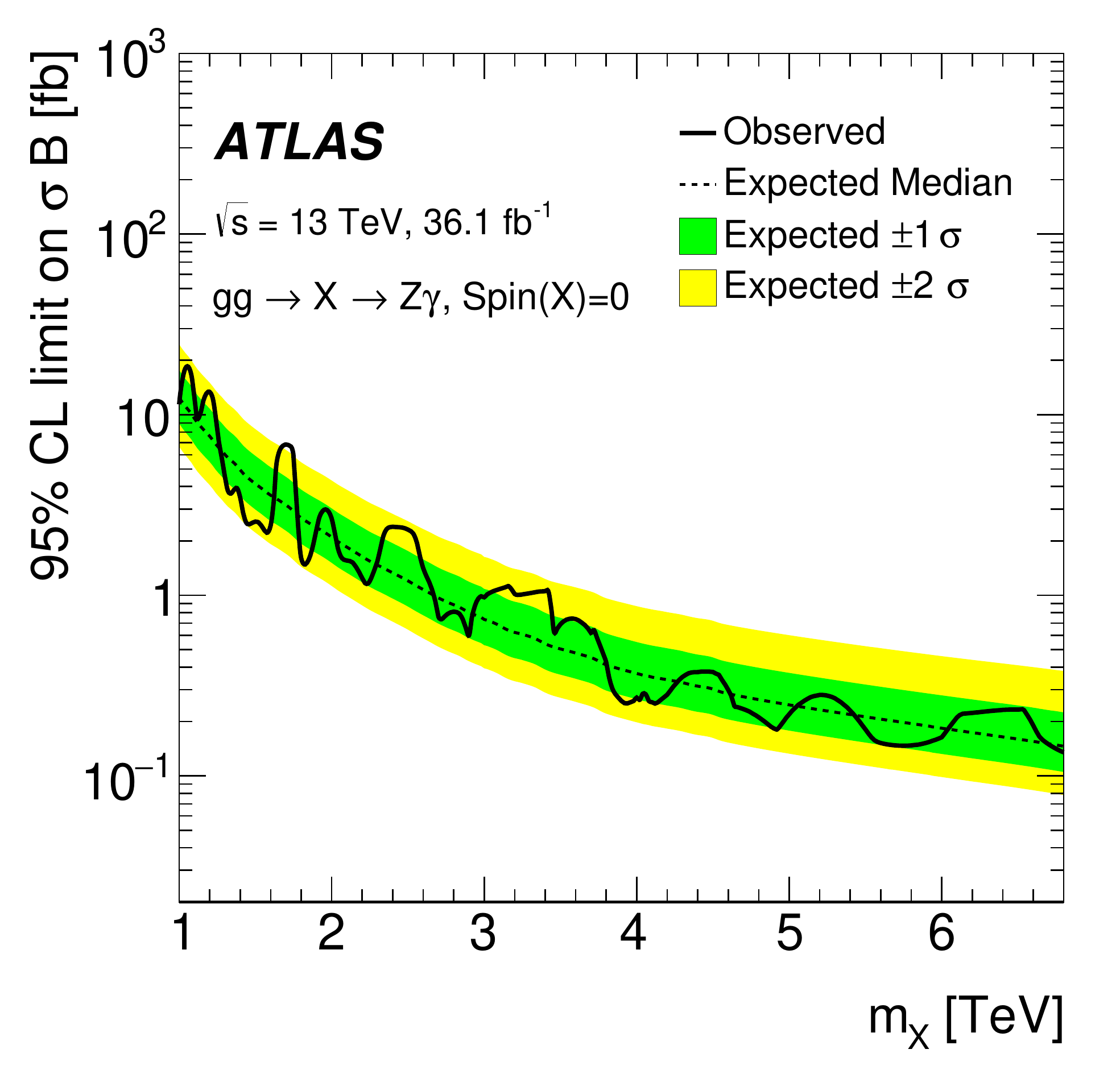}}
    \subfigure[]{\includegraphics[width=0.48\columnwidth]{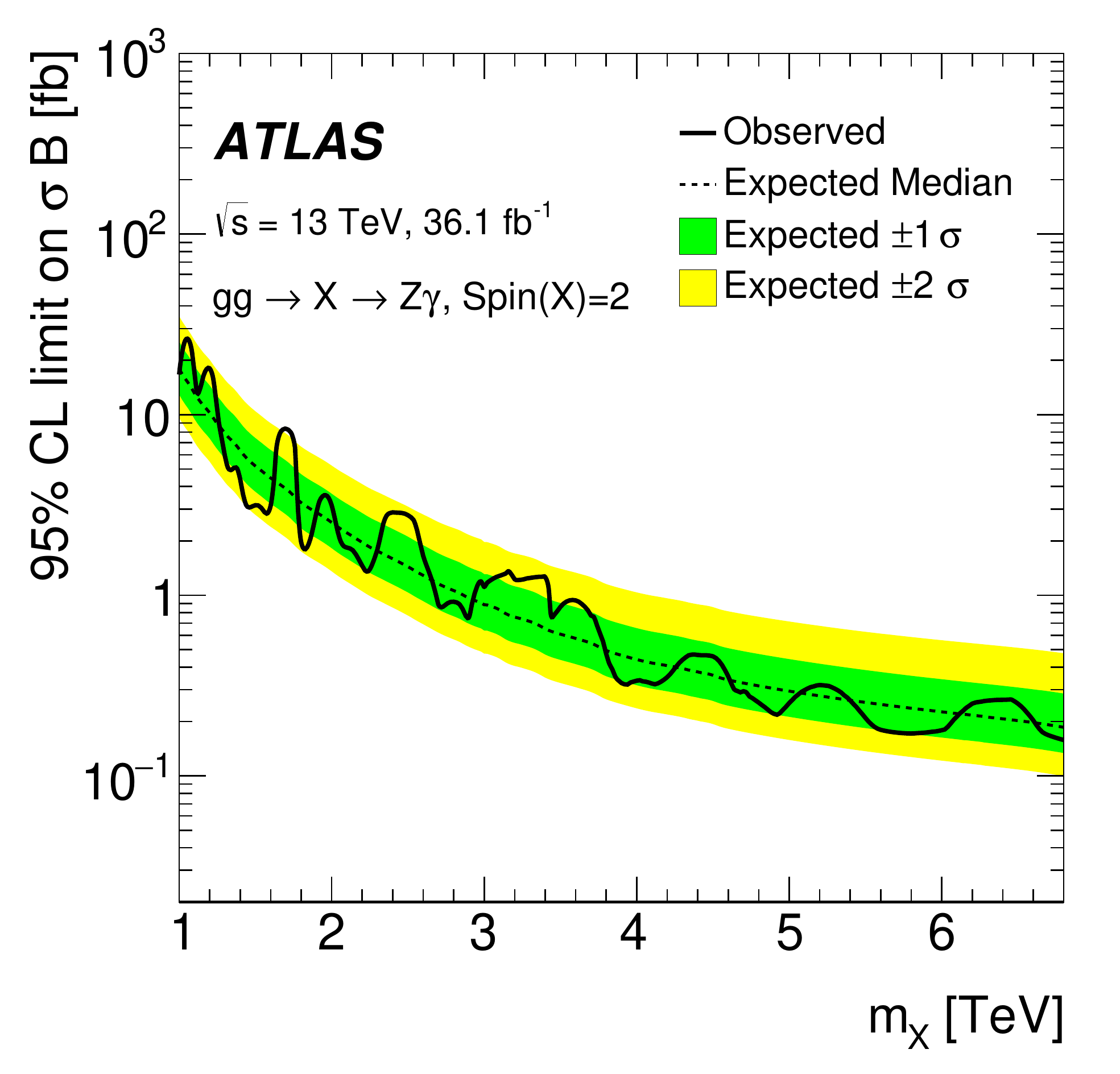}}
    \subfigure[]{\includegraphics[width=0.48\columnwidth]{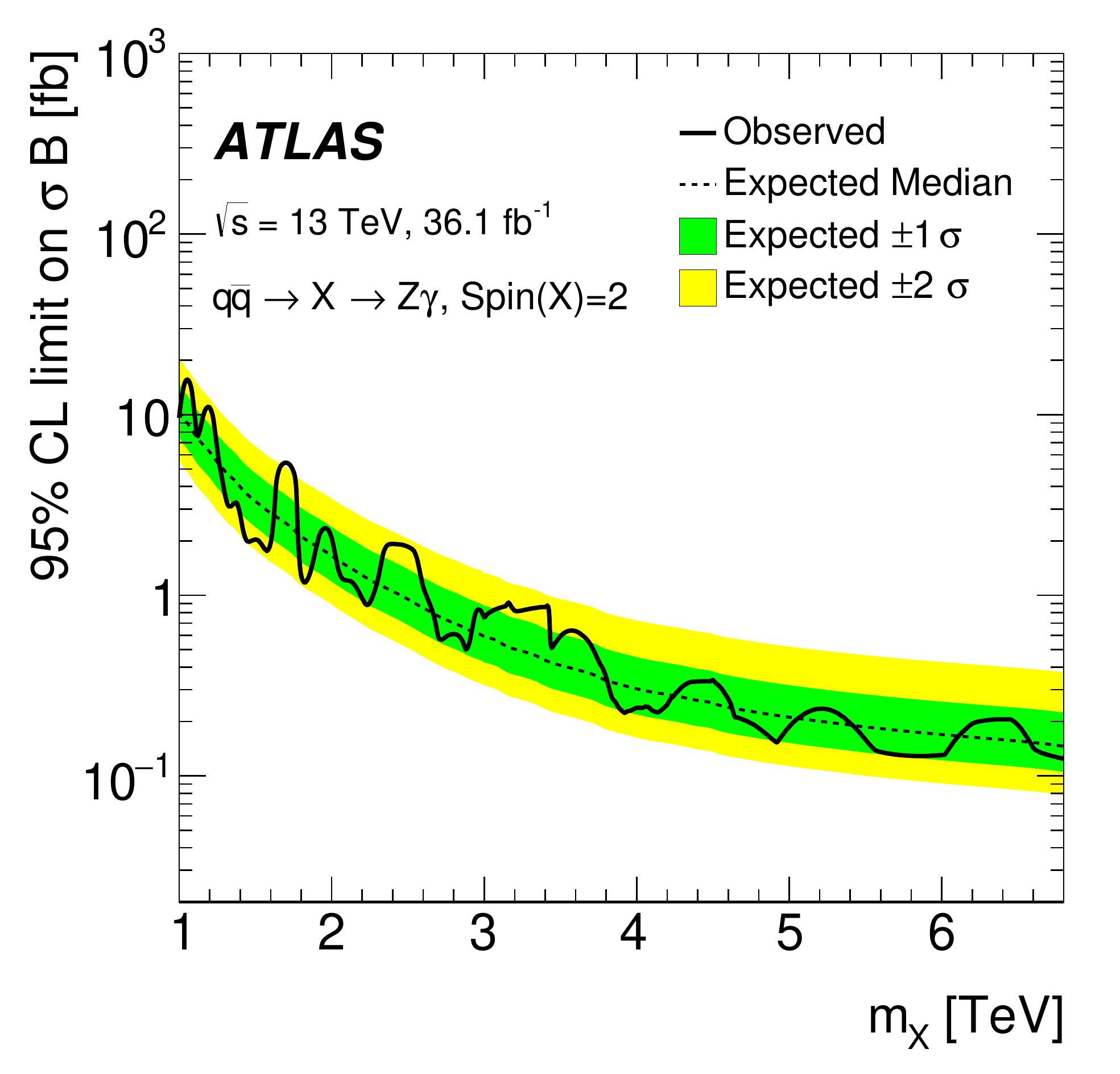}}
      \caption{The 95\% CL observed (solid line) and expected (dashed line) upper limits on $\sigma B$ for a resonance with (a) spin-0, produced by gluon--gluon fusion,
      (b) spin-2, produced by gluon--gluon fusion, and (c) spin-2, produced by quark--antiquark annihilation, decaying to $Z\gamma$ as a function of the resonance mass.
       The inner and outer bands give the $\pm 1$ and $\pm 2$ standard deviations of expected limits.
       The local deviations of the observed upper limits from the expected ones on $\sigma B$ are a result of small deviations in the data from the best-fitting background-only model.
    }
    \label{fig:limitsZ}
  \end{center}
\end{figure}

\begin{figure}[!h]
  \begin{center}
    \includegraphics[width=0.48\columnwidth]{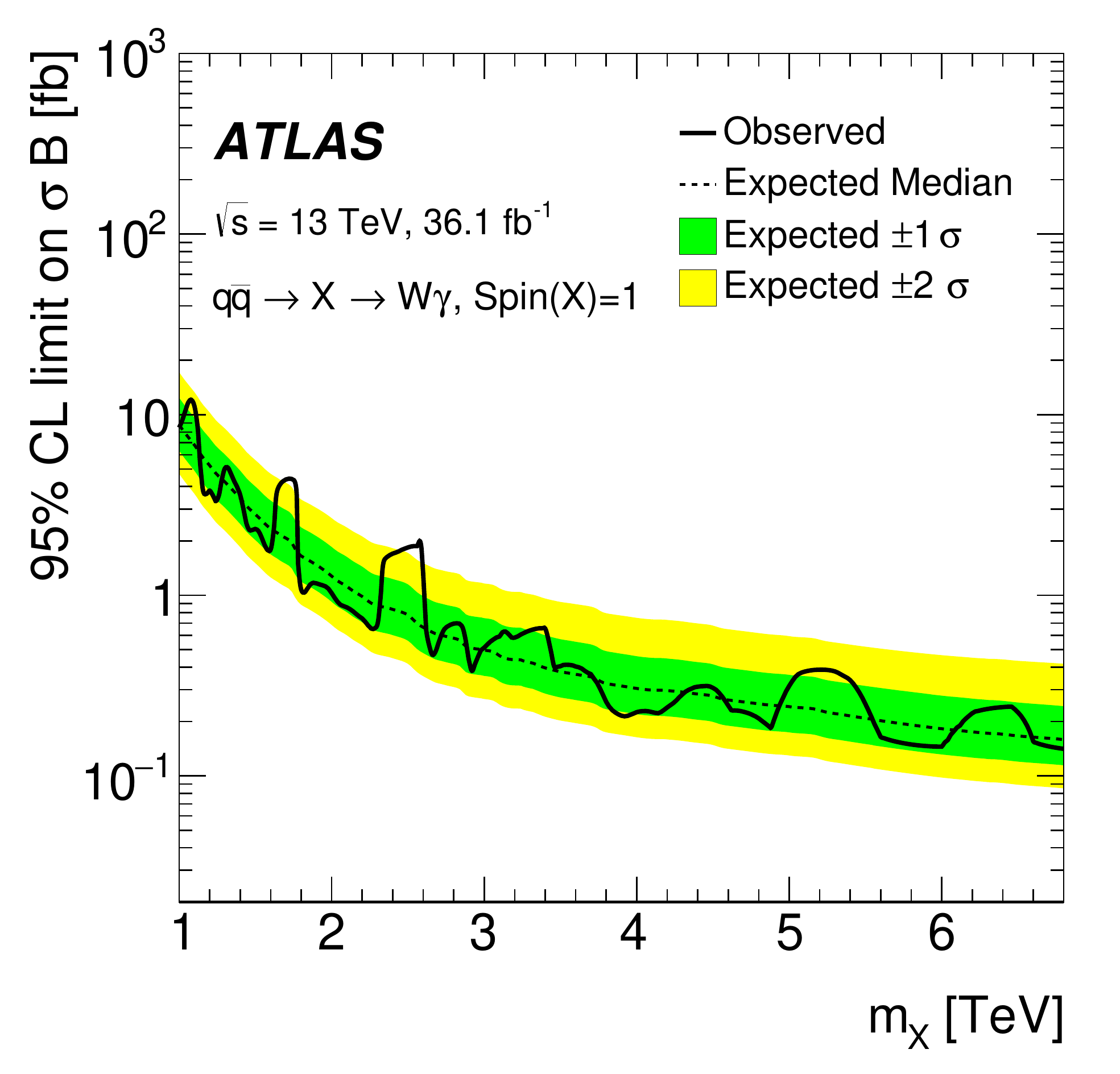}
    \caption{The 95\% CL observed (solid line) and expected (dashed line) upper limits on $\sigma B$ for a spin-1 resonance decaying to $W\gamma$ as a function of the resonance mass.
       The inner and outer bands give the $\pm 1$ and $\pm 2$ standard deviations of expected limits.
       The local deviations of the observed upper limits from the expected ones on $\sigma B$ are a result of small deviations in the data from the best-fitting background-only model.
    }
    \label{fig:limitsW}
  \end{center}
\end{figure}

\begin{figure}[!h]
  \begin{center}
    \includegraphics[width=0.48\columnwidth]{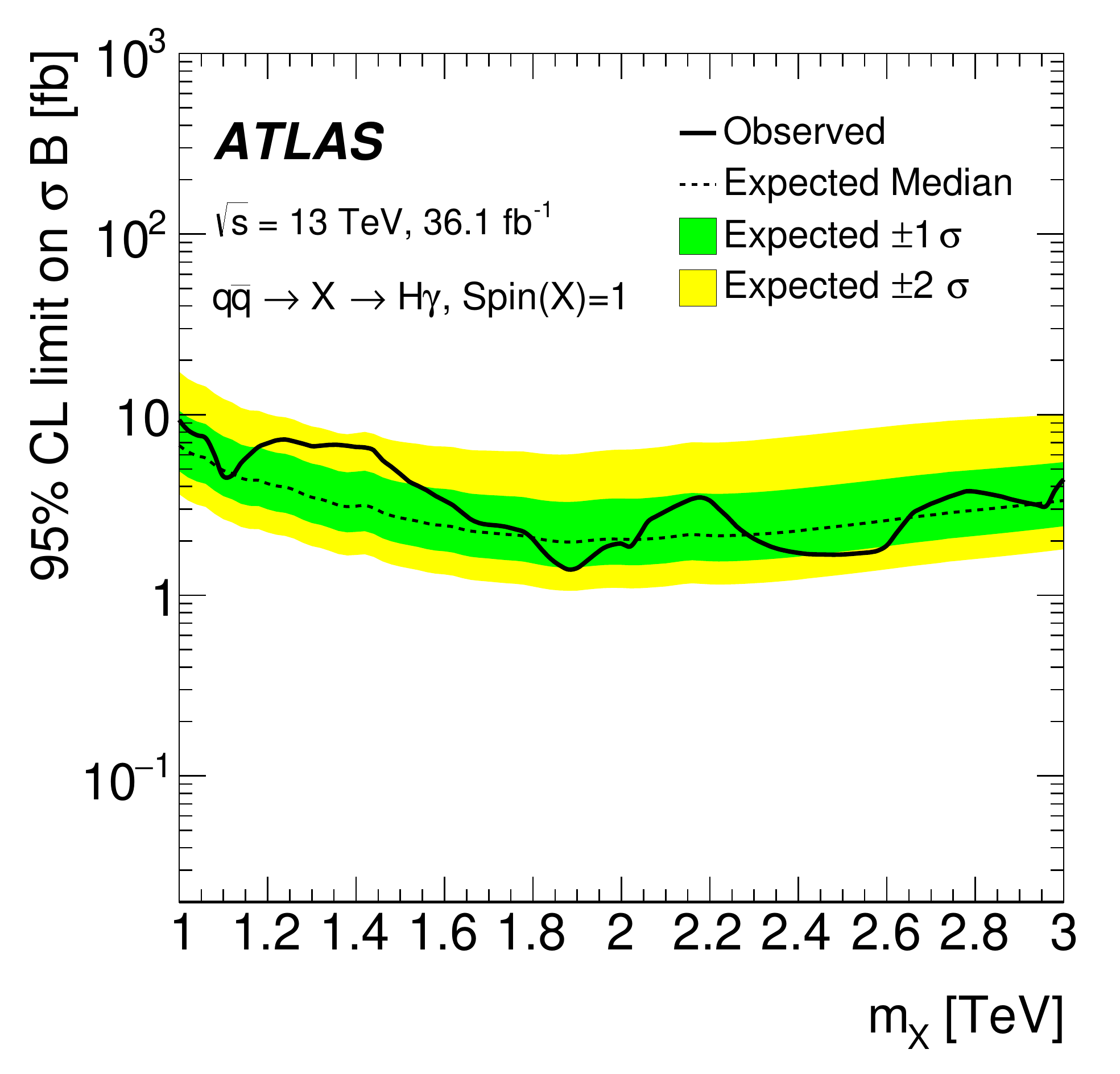}
    \caption{The 95\% CL observed (solid line) and expected (dashed line) upper limits on $\sigma B$ for a spin-1 resonance decaying to $H\gamma$ as a function of the resonance mass.
       The inner and outer bands give the $\pm 1$ and $\pm 2$ standard deviations of expected limits.
       The local deviations of the observed upper limits from the expected ones on $\sigma B$ are a result of small deviations in the data from the best-fitting background-only model.
    }
    \label{fig:limitsH}
  \end{center}
\end{figure}

\FloatBarrier
\section{Conclusion}
\label{sec:conclusion}

Results are presented from a search for heavy resonances decaying to $Z\gamma$,
$W\gamma$ or $H\gamma$ final states using 36.1 \ifb{} of $\sqrt{s}$ = 13 \TeV{} $pp$ collision data collected by the
ATLAS experiment during the 2015 and 2016 at the LHC run periods. The search sensitivity
is enhanced for high-mass resonances by selecting hadronic decays of the $Z$
and $W$ bosons, and the $b\bar{b}$ decay of the Higgs boson, identified as large-radius jets.
Distributions of the invariant mass of photon--jet
pairs are used to search for resonances in the mass range from 1.0 to 6.8~TeV{} for decays to $Z\gamma$ and $W\gamma$, and between 1.0 and 3.0~\TeV{} for decays to $H\gamma$.
No evidence for new resonances is found and limits are set based on assumptions on the spin and production model of the resonance.

The 95$\%$ confidence level upper limits on the resonance production cross section times
decay branching fraction to $Z\gamma$ and $W\gamma$ final states vary from
about 10~fb to 0.1~fb for masses from 1.0 to 6.8~\TeV{}, with individual studies
carried out for resonances with spin  0, 1, and 2 produced via gluon--gluon
fusion and $q\bar{q}$ annihilation. For the spin-1 $X \to$ $H\gamma$ search the limits vary from about 10~fb to 4~fb for resonance masses between 1.0 and 3.0~\TeV{}.
These results set the first limits on the production of
$H\gamma$ resonances, and this search covers a wider mass range and has a broader scope than previous
searches for heavy resonances decaying to $Z\gamma$ and
$W\gamma$ final states.

\section*{Acknowledgements}

We thank CERN for the very successful operation of the LHC, as well as the
support staff from our institutions without whom ATLAS could not be
operated efficiently.

We acknowledge the support of ANPCyT, Argentina; YerPhI, Armenia; ARC, Australia; BMWFW and FWF, Austria; ANAS, Azerbaijan; SSTC, Belarus; CNPq and FAPESP, Brazil; NSERC, NRC and CFI, Canada; CERN; CONICYT, Chile; CAS, MOST and NSFC, China; COLCIENCIAS, Colombia; MSMT CR, MPO CR and VSC CR, Czech Republic; DNRF and DNSRC, Denmark; IN2P3-CNRS, CEA-DRF/IRFU, France; SRNSFG, Georgia; BMBF, HGF, and MPG, Germany; GSRT, Greece; RGC, Hong Kong SAR, China; ISF, I-CORE and Benoziyo Center, Israel; INFN, Italy; MEXT and JSPS, Japan; CNRST, Morocco; NWO, Netherlands; RCN, Norway; MNiSW and NCN, Poland; FCT, Portugal; MNE/IFA, Romania; MES of Russia and NRC KI, Russian Federation; JINR; MESTD, Serbia; MSSR, Slovakia; ARRS and MIZ\v{S}, Slovenia; DST/NRF, South Africa; MINECO, Spain; SRC and Wallenberg Foundation, Sweden; SERI, SNSF and Cantons of Bern and Geneva, Switzerland; MOST, Taiwan; TAEK, Turkey; STFC, United Kingdom; DOE and NSF, United States of America. In addition, individual groups and members have received support from BCKDF, the Canada Council, CANARIE, CRC, Compute Canada, FQRNT, and the Ontario Innovation Trust, Canada; EPLANET, ERC, ERDF, FP7, Horizon 2020 and Marie Sk{\l}odowska-Curie Actions, European Union; Investissements d'Avenir Labex and Idex, ANR, R{\'e}gion Auvergne and Fondation Partager le Savoir, France; DFG and AvH Foundation, Germany; Herakleitos, Thales and Aristeia programmes co-financed by EU-ESF and the Greek NSRF; BSF, GIF and Minerva, Israel; BRF, Norway; CERCA Programme Generalitat de Catalunya, Generalitat Valenciana, Spain; the Royal Society and Leverhulme Trust, United Kingdom.

The crucial computing support from all WLCG partners is acknowledged gratefully, in particular from CERN, the ATLAS Tier-1 facilities at TRIUMF (Canada), NDGF (Denmark, Norway, Sweden), CC-IN2P3 (France), KIT/GridKA (Germany), INFN-CNAF (Italy), NL-T1 (Netherlands), PIC (Spain), ASGC (Taiwan), RAL (UK) and BNL (USA), the Tier-2 facilities worldwide and large non-WLCG resource providers. Major contributors of computing resources are listed in Ref.~\cite{ATL-GEN-PUB-2016-002}.

\printbibliography


\clearpage

 
\begin{flushleft}
{\Large The ATLAS Collaboration}

\bigskip

M.~Aaboud$^\textrm{\scriptsize 34d}$,    
G.~Aad$^\textrm{\scriptsize 99}$,    
B.~Abbott$^\textrm{\scriptsize 124}$,    
O.~Abdinov$^\textrm{\scriptsize 13,*}$,    
B.~Abeloos$^\textrm{\scriptsize 128}$,    
S.H.~Abidi$^\textrm{\scriptsize 165}$,    
O.S.~AbouZeid$^\textrm{\scriptsize 143}$,    
N.L.~Abraham$^\textrm{\scriptsize 153}$,    
H.~Abramowicz$^\textrm{\scriptsize 159}$,    
H.~Abreu$^\textrm{\scriptsize 158}$
% ...
\bigskip
\\

% ...
$^{13}$Institute of Physics, Azerbaijan Academy of Sciences, Baku; Azerbaijan.\\
% ...
$^{34}$$^{(a)}$Facult\'e des Sciences Ain Chock, R\'eseau Universitaire de Physique des Hautes Energies - Universit\'e Hassan II, Casablanca;$^{(b)}$Centre National de l'Energie des Sciences Techniques Nucleaires (CNESTEN), Rabat;$^{(c)}$Facult\'e des Sciences Semlalia, Universit\'e Cadi Ayyad, LPHEA-Marrakech;$^{(d)}$Facult\'e des Sciences, Universit\'e Mohamed Premier and LPTPM, Oujda;$^{(e)}$Facult\'e des sciences, Universit\'e Mohammed V, Rabat; Morocco.\\
% ...
$^{99}$CPPM, Aix-Marseille Universit\'e, CNRS/IN2P3, Marseille; France.\\
% ...
$^{124}$Homer L. Dodge Department of Physics and Astronomy, University of Oklahoma, Norman OK; United States of America.\\
% ...
$^{128}$LAL, Universit\'e Paris-Sud, CNRS/IN2P3, Universit\'e Paris-Saclay, Orsay; France.\\
% ...
$^{143}$Santa Cruz Institute for Particle Physics, University of California Santa Cruz, Santa Cruz CA; United States of America.\\
% ...
$^{153}$Department of Physics and Astronomy, University of Sussex, Brighton; United Kingdom.\\
% ...
$^{158}$Department of Physics, Technion, Israel Institute of Technology, Haifa; Israel.\\
$^{159}$Raymond and Beverly Sackler School of Physics and Astronomy, Tel Aviv University, Tel Aviv; Israel.\\
% ...
$^{165}$Department of Physics, University of Toronto, Toronto ON; Canada.\\
% ...
$^{a}$ Also at  Department of Physics, University of Malaya, Kuala Lumpur; Malaysia.\\
$^{b}$ Also at Borough of Manhattan Community College, City University of New York, NY; United States of America.\\
$^{c}$ Also at Centre for High Performance Computing, CSIR Campus, Rosebank, Cape Town; South Africa.\\
$^{d}$ Also at CERN, Geneva; Switzerland.\\
$^{e}$ Also at CPPM, Aix-Marseille Universit\'e, CNRS/IN2P3, Marseille; France.\\
% ...
$^{*}$ Deceased

\end{flushleft}


\end{document}